\documentclass[red,11pt,a4paper]{article}
\usepackage{cite}

\usepackage{amsmath}
\usepackage{amscd}
\usepackage{amssymb}
\usepackage{graphicx}
\usepackage{latexsym}
\usepackage{color}
\usepackage{floatrow}
\newfloatcommand{capbtabbox}{table}[][\FBwidth]

\usepackage{blindtext}

\newcommand{\lap}{\Delta}

\newcommand{\vy}{Y_{k}}
\newcommand{\vf}{{\vc F}}

\newcommand{\vr}{\varrho}

\newcommand{\vt}{\vartheta}

\newcommand{\vu}{\vc{u}}

\newcommand{\vc}[1]{{\bf #1}}

\newcommand{\Grad}{\nabla}

\newcommand{\pt}{\partial_{t}}

\newcommand{\dx}{{\rm d} {x}}
\newcommand{\dt}{{\rm d} t }

\newcommand{\dxdt}{\dx \ \dt}

\newcommand{\lr}[1]{\left( #1 \right)}

\newcommand{\eq}[1]{\begin{equation}
\begin{split}
#1
\end{split}
\end{equation}}
\newcommand{\eqh}[1]{\begin{equation*}
\begin{split}
#1
\end{split}
\end{equation*}}
\newcommand{\ep}{\varepsilon}

\newtheorem{thm}{Theorem}

\newtheorem{prop}[thm]{Proposition}

\title{Particle interactions mediated by dynamical networks: assessment of
macroscopic descriptions}
\author{J. Barr\'e$^{1,2}$, J. A. Carrillo$^3$, P. Degond$^3$, D. Peurichard$^{4}$, E. Zatorska$^3$}
\date{}

\begin{document}
\maketitle

\centerline{1.  Laboratoire MAPMO, CNRS, UMR 7349, F\'ed\'eration Denis Poisson,}
\centerline{ FR 2964, Universit\'e d'Orl\'eans, B.P. 6759, 45067 Orl\'eans cedex 2, France.}
\bigskip
\centerline{2. Institut Universitaire de France, Paris, France.}
 \bigskip
\centerline{3. Department of Mathematics, Imperial College London, }
\centerline{London SW7 2AZ, United Kingdom.}
\bigskip
\centerline{4. Faculty of Mathematics, University of Vienna,}
\centerline{Oskar-Morgenstern Platz 1, 1090 Vienna, Austria.}

\begin{abstract}
We provide a numerical study of the macroscopic model of \cite{BDZ2016} derived from an agent-based model for a system of particles interacting through a dynamical network of links. Assuming that the network remodelling process is very fast, the macroscopic model takes the form of a single aggregation diffusion equation for the density of particles. The theoretical study of the macroscopic model gives precise criteria for the phase transitions of the steady states, and in the 1-dimensional case, we show numerically that the stationary solutions of the microscopic model undergo the same phase transitions and bifurcation types as the macroscopic model. In the 2-dimensional case, we show that the numerical simulations of the macroscopic model are in excellent agreement with the predicted
theoretical values. This study provides a partial validation of the formal
derivation of the macroscopic model from a microscopic formulation
and shows that the former is
a consistent approximation of an underlying particle dynamics,
making it a powerful tool for the modelling of dynamical networks
at a large scale.
\end{abstract}



\section{Introduction}

Complex networks are of significant interest in many fields of
life and social sciences. These systems are composed of a large
number of agents interacting through local interactions, and
self-organizing to reach large-scale functional structures.
Examples of systems involving highly dynamical networks include
neural networks, biological fiber networks such as connective tissues, vascular or neural networks, ant trails, polymers, economic
interactions etc \cite{Boissard_etal_JMB13, Mog1999,Dido2006,Broe2010}. These networks often
offer great plasticity by their ability to break and reform
connections, giving to the system the ability to change shape and
adapt to different situations \cite{Boissard_etal_JMB13, Chaud2007}. Because of their
paramount importance in biological functions or social
organizations, understanding the properties of such complex
systems is of great interest. However, they are challenging to
model due to the large amount of components and interactions
(chemical, biological, social etc). Due to their simplicity and
flexibility, individual based models are a natural framework to
study complex systems. They describe the behavior of each agent
and its interaction with the surrounding agents over time,
offering a description of the system at the microscopic scale (see e.g. 
\cite{BDZ2016, Boissard_etal_JMB13, Degond_etal_M3AS16}).
However, these models are computationally expensive and are not
suited for the study of large systems. To study the systems at a
macroscopic scale, mean-field or continuous models are often
preferred. These last models describe the evolution in time of
averaged quantities such as agent density, mean orientation etc.
As a drawback, these last models lose the information at the
individual level. In order to overcome this weakness of the
continuous models, a possible route is to derive a macroscopic
model from an agent-based formulation and to compare the obtained
systems, as was done in e.g. \cite{BDZ2016, Boissard_etal_JMB13, Degond_etal_M3AS16} for particle interactions mediated by dynamical networks.

A first step in this direction has been made in \cite{BDZ2016},
following the earlier work \cite{Degond_etal_M3AS16}. In this work, the derivation of a macroscopic model
for particles interacting through a dynamical network of links is performed.
The microscopic model describes the evolution in time of point
particles which interact with their close neighbors via local
cross-links modelled by springs that are randomly created and
destructed. In the mean field limit, assuming large number of
particles and links as well as propagation of chaos, the
corresponding kinetic system consists of two equations: for the
individual particle distribution function, and for the link
densities. The link density distribution provides a statistical
description of the network connectivity which turns out to be
quite flexible and easily generalizable to other types of complex
networks.

In the large scale limit and in the regime where link
creation/destruction frequency is very large, it was shown in
\cite{BDZ2016}, following \cite{Degond_etal_M3AS16}, that the link density distribution becomes a local
function of the particle distribution density. The latter evolves
on the slow time scale through an aggregation-diffusion equation.
Such equations are encountered in many physical systems featuring
collective behavior of animals, chemotaxis models, etc
\cite{TBL,BDP,CFTV10,Goles11,Kolo2013} and references
therein. The difference between this macroscopic model and the
aggregation-diffusion equations studied in the literature
\cite{CMV03,TBL,Berto2009} lies in the fact that the interaction
potential has compact support. As a result, this model has a rich
behavior such as metastability in the case of the whole space
\cite{BFH14,EK16} and exhibits phase transitions in the periodic
setting as functions of the diffusion coefficient, the interaction
range of the potential and the links equilibrium length \cite{BDZ2016}. By
performing the weakly nonlinear stability analysis of the
spatially homogeneous steady states, it is possible
to characterize the type of bifurcations appearing at the
instability onset \cite{BDZ2016}. We refer to \cite{Barbaro_Degond_DCDSB14, ChPa,DFL15,BCCD16} for
related collective dynamics problems showing phase transitions.

If numerous macroscopic models for dynamical networks have been
proposed in the literature, most of them are based on
phenomenological considerations and very few have been linked to
an agent-based dynamics. On the contrary, the macroscopic model
proposed in \cite{BDZ2016} and its precursor \cite{Degond_etal_M3AS16} have been derived via a formal mean
field limit from an underlying particle dynamics (see also \cite{Degond_etal_MMS14}). However, because
the derivation performed in \cite{BDZ2016} is still formal, its numerical validation as the
limit of the microscopic model as well as the persistence of the
phase transitions at the micro and macroscopic level as predicted
by the weakly nonlinear analysis in \cite{BDZ2016} need to be assessed. This is the goal of the present work.

More precisely, we show that the macroscopic model indeed provides
a consistent approximation of the underlying agent-based model for
dynamical networks, by confronting numerical simulations of both
the micro- and macro- models. Moreover, we numerically check that
the microscopic system undergoes in 1-dimensional a phase transition depicted
by the values obtained for the limiting macroscopic
aggregation-diffusion equation. Furthermore, we numerically
validate the weakly nonlinear analysis in \cite{BDZ2016} for the
type of bifurcation in the 2-dimensional setting, where simulations for the microscopic
model are prohibitively expensive.

The paper is organized as follows. In Section \ref{Sec:deriv}, we
present the microscopic model and sketch the derivation of the
kinetic and macroscopic models from the agent-based formulation.
In Section \ref{Sec:macro}, we focus on the 1-dimensional case: we first
summarize the theoretical results on the stability of homogeneous
steady states of the macroscopic model from \cite{BDZ2016}, and
show that both the macroscopic and microscopic simulations are in
good agreement with the theoretical predictions made by nonlinear
analysis of the macroscopic model. We then compare the profiles of
the steady states between the microscopic and macroscopic
simulations, and show that the two formulations are in very good
agreement, also in terms of phase transitions. Finally, in Section
\ref{Sec:macro2D} we provide a numerical study of the 2-dimensional case for
the macroscopic model. The 2-dimensional numerical simulations on the
macroscopic model are able to numerically capture the subcritical
and supercritical transitions as predicted theoretically. Because
of the computational cost of the microscopic model, the
macroscopic model is not only very competitive and efficient in
order to detect phase transitions but also it is almost the only
feasible choice showing the main advantage of the limiting kinetic
procedure.

%
%

\section{Derivation of the macroscopic model}\label{Sec:deriv}

\subsection{Microscopic model}\label{SSec:micro}

\begin{figure}[h!]
\includegraphics[scale=0.2]{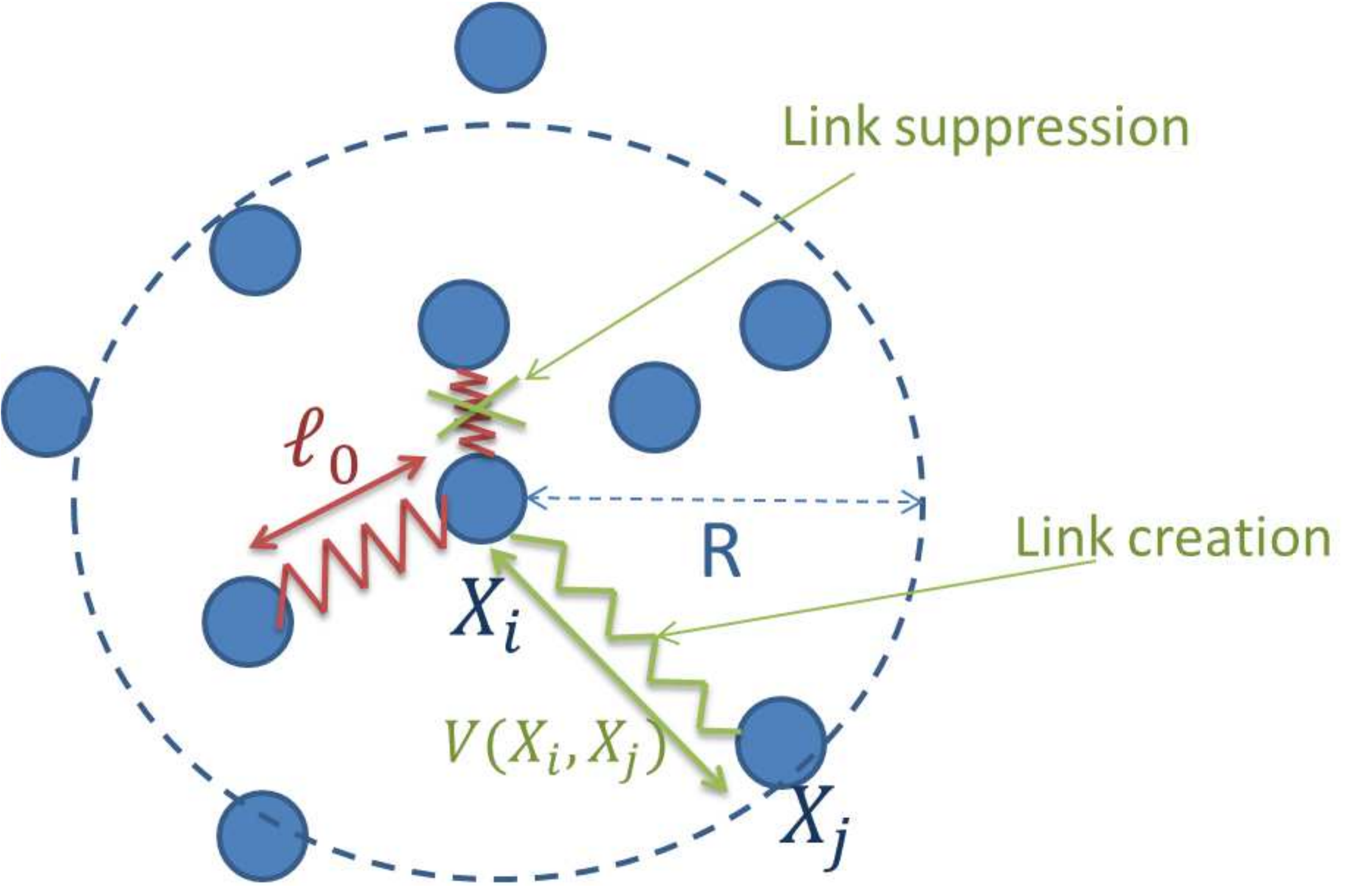}
\caption{Particles interacting through a network of links seen as springs of equilibrium length $l$. The detection zone for linking to close neighbours is a disk of radius $R$. Link suppression/creation is supposed to be random in time. \label{Scheme}}
\end{figure}

The 2-dimensional microscopic model features $N$ particles located at points $X_i \in \Omega, i\in [1,N]$ linking/unlinking -dynamically in time- to their neighbors which are located in a ball of radius $R$ from their center. The link creation and suppression are supposed to follow Poisson processes in time, of frequencies $\nu_f^N$ and $\nu^N_d$ respectively (see Fig.\ref{Scheme}).

 Each link is supposed to act as a spring by generating a pairwise potential
\eq{\label{Vform}
\tilde V(X_i,X_j) = U(|X_i-X_j|) = \frac{\kappa}{2}(|X_i - X_j| - \ell)^2,
}
\noindent where $\kappa$ is the intensity of the spring force and $\ell$ the equilibrium length of the spring. We define the total energy of the system $W$ related to the maintenance of the links:
\begin{equation}\label{W}
W = \sum_{k=1}^K \tilde V(X_{i(k)},X_{j(k)}),
\end{equation}
\noindent where $i(k),j(k)$ denote the indexes of particles connected by the link $k$. Particle motion between two linking/unlinking events is then supposed to occur in the steepest descent direction to this energy, in the so-called overdamped regime:
\begin{equation}\label{Micro_Eq}
dX_i = - \mu \nabla_{X_i} W dt + \sqrt{2D}dB_i,
\end{equation}
\noindent for $i \in [1,N]$ and where $B_i$ is a 2-dimensional Brownian motion $B_i = (B_i^1,B_i^2)$ with diffusion coefficient $D>0$ and $\mu>0$ is the mobility coefficient.

\subsection{Kinetic model}
To perform the mean-field limit, following \cite{BDZ2016} and \cite{Degond_etal_M3AS16}, we define the one particle distribution of the $N$ particles, $f^N(x,t)$, and the link distribution of the $K$ links, $g^K(x_1,x_2,t)$. Postulating the existence of the following limits:
$$
f(x,t) = \underset{N \rightarrow \infty}{\lim}{f^N}, \quad g(x_1,x_2,t) = \underset{K \rightarrow \infty}{\lim}{g^K}, $$
$$\nu_f = \underset{N \rightarrow \infty}{\lim}{\nu_f^N(N-1)}, \quad \nu_d = \underset{N \rightarrow \infty}{\lim}{\nu_d^N}, \quad\xi = \underset{K,N \rightarrow \infty}{\lim}{\frac{K}{N}}
$$
\noindent the kinetic system reads:
\begin{equation*}
\partial_t f(x,t) = D \Delta_x f(x,t) + 2\mu \xi \nabla_x \cdot F(x,t)
\end{equation*}
\begin{align*}
\partial_t g(x_1,x_2,t) =& D (\Delta_{x_1} g + \Delta_{x_2} g) + 2\mu \xi \left[\nabla_{x_1} \cdot \left(\frac{g(x_1,x_2)}{f(x_1)} F(x_1,t) \right)+ \nabla_{x_2} \cdot \left(\frac{g(x_1,x_2)}{f(x_2)} F(x_2,t)\right) \right] \nonumber\\
&+ \frac{\nu_f}{2\xi} f(x_1,t)f(x_2,t) \chi_{|x_1-x_2|\leq R} - \nu_d g(x_1,x_2,t),
\end{align*}
\noindent where we have postulated that the distribution of pairs of particles reduces to $f(x_1,t)f(x_2,t)$, and
$$
F(x,t) = \int g(x,y,t)\nabla_{x_1} \tilde V(x,y) dxdy.
$$
We refer the reader to \cite{BDZ2016} for details on the mean-field limit.

\subsection{Scaling and macroscopic model}\label{Ss:scaling}
In this paper, the space and time scales are chosen such that $\mu = 1$
and the variables are scaled such that:
$$
\tilde{x} = \varepsilon^{1/2} x,\quad \tilde{t} = \varepsilon t, \quad f^\varepsilon(\tilde{x},\tilde{t}) = \varepsilon^{-1} f(x,t), \quad g^\varepsilon(\tilde{x}_1,\tilde{x}_2,\tilde{t}) = \varepsilon^{-2} g(x_1,x_2,t).
$$
The spring force $\kappa$ is supposed to be small, i.e $\tilde{\kappa} = \varepsilon^{-1} \kappa$, the noise $D$ is supposed to be of order 1 and the typical spring length $\ell$ and particle detection distance $R$ are supposed to scale as the space variable, i.e $\tilde \ell = \varepsilon^{1/2} \ell$, $\tilde{R} = \varepsilon^{1/2} R$. Finally, the main scaling assumption is to consider that the processes of linking and unlinking are very fast, i.e $\tilde{\nu}_f = \varepsilon^2 \nu_f, \tilde{\nu}_d = \varepsilon^2 \nu_d$. For the sake of simplicity, we will consider in this paper that $\frac{\tilde \nu_f}{\tilde \nu_d} = 1$, and $\tilde{\kappa} = 2$.

For such a scaling, it is shown in \cite{BDZ2016} that in the limit $\varepsilon \rightarrow 0$, if we suppose $(f^\varepsilon,g^\varepsilon) \rightarrow_{\varepsilon \rightarrow 0} (f, g)$, then:
 \begin{subequations}\label{Macro_sys}
\eq{
\partial_t f = D \Delta_x f + \nabla_x \cdot \big(f \; (\nabla_x V\ast f) \big) \label{Macro}}
\eq{
g(x,y,t) = \frac{\nu_f}{2\xi \nu_d} f(x,t)f(y,t) \chi_{|x-y|\leq R},\label{Macro_g}}
\end{subequations}
\noindent for some compactly supported potential $V$ such that:
$$
\nabla_x V = U'(|x|) \chi_{|x|\leq R} \frac{x}{|x|}  
$$
In this paper, we take $\kappa=2$ in~\eqref{Vform}, hence $V$ has the form:
\begin{equation}{\label{TVH}
 V(x)=
\left\{
\begin{array}{lll}
(|x|-\ell)^2-(R-\ell)^2,& \mbox{for}&|x|< R,\\
0& \mbox{for}&|x|\geq R.
\end{array}
\right.
}
\end{equation}

In the following, we aim to study theoretically and numerically both the macroscopic model given by Eqs. \eqref{Macro_sys}, and the corresponding microscopic formulation given by Eq. \eqref{Micro_Eq} and rescaled with the scaling introduced in this section. We first focus on the 1-dimensional case and we show that the numerical solutions behave as theoretically predicted, and that we obtain -- numerically -- a very good agreement between the micro- and macro- formulations.

%
%
\section{Analysis of the macroscopic model in the 1-dimensional case}\label{Sec:macro}

\subsection{Theoretical results}

In this section, we apply the results of \cite{BDZ2016} to the 1 dimensional periodic domain $[-L,L]$,  to study the stability of stationary solutions of the macroscopic model given by Eq.~\eqref{Macro}.

\subsubsection{Identification of the stability region}\label{Swc:3111}

We first linearize equation~\eqref{Macro} around the constant steady state $\rho^*=\frac{1}{2L}$, so that the total mass is equal to 1, we denote the perturbation by $\rho$, so we have $f=\rho^*+\rho$, that satisfies

\eq{\label{start_lin}
\partial_{t}  \rho={D\lap_{x}  \rho}+\rho^*\lap (V\ast \rho),
}
where $V$ is given by~\eqref{TVH}. We will further decompose $f$ into its Fourier modes
\eqh{
\rho(x)=\sum_{k\in \mathbb{Z}}\hat \rho_{k}e_{k},\quad \mbox{where}\quad  e_{k}=\exp{\left[i\pi \frac{kx}{L}\right]}
}
the Fourier transform is given by
\eqh{
\hat \rho_{k}=\frac{1}{2L}\int_{-L}^{L} \rho(x)e_{-k}\, dx.}
Applying the Fourier transform to~\eqref{start_lin}, a straightforward computation gives
\eq{\label{fF}
\pt\hat \rho_k=-\lr{\frac{\pi k}{L}}^2\lr{D+\hat V_k}\hat \rho_k,
}
where the Fourier modes of the potential $V$ are given by
\eq{\label{Vk}
\hat V_k&=\frac{2R^3 }{L}\lr{-\frac{\sin(z_k)}{z_k^3}+(1-\alpha)\frac{\cos(z_k)}{z_k^2}+\frac{\alpha}{z_k^2}}.
}
Here, we denoted
$$\alpha=\frac{\ell}{R},\qquad z_k=\frac{\pi R |k|}{L}.$$
Therefore, the stability of the constant steady state will be
ensured if the coefficient in front of $\hat\rho_k$ on the r.h.s. of~\eqref{fF} has a non-positive real part
for $k=1$. This condition is related to the H-stable/catastrophic behavior of interaction potentials that characterizes the existence of global minimizers of the total potential energy as recently shown in \cite{CaCaPa,SiSlTo15}.

\subsubsection{Characterization of  the bifurcation type}

As shown in \cite{BDZ2016}, it is possible to distinguish two types of bifurcation as functions of the model parameters. Indeed, if we define:
 \begin{subequations}
\eq{
\lambda=\lambda_{\pm 1}=-\frac{\pi^2}{L^2}\lr{D+\hat{ V}_{1}}, \label{lambda1}}
\eq{\lambda_{k}=-\frac{\pi^2k^2}{L^2}\lr{D+\hat{V}_{k}}, \label{lambdak}}
 \end{subequations}
\noindent we have the following proposition (see \cite{BDZ2016}):
\begin{prop}\label{bifurc}
Assume that $\lambda>0$ and $\lambda_{k}<0, \; \forall k \neq \pm 1$. Then:
\begin{itemize}
\item if $2\hat{V}_{2} - \hat{V}_{-1}>0$, the steady state exhibits a supercritical bifurcation;
\item if $2\hat{V}_{2} - \hat{V}_{-1}<0$, the steady state exhibits a subcritical bifurcation.
\end{itemize}
\end{prop}
Note that the above criterion only involves the potential but does not involve the parameter $D$, it only restricts the values of $\alpha$ or $\ell$.

\subsection{Numerical results}
\subsubsection{Choice of numerical parameters}
In the linearized equation \eqref{start_lin}, there are four parameters that may vary: $D$, $\ell$, $R$ and $L$.
In this part of the paper, we focus on the case where the potential is of comparable range $R$ to the size of the domain $L$, and fix the value of the following parameters:
$$L=3\quad \mbox{and} \quad R=0.75,$$
therefore $z_1=\frac{\pi}{4}$. Using \eqref{fF}, and the discussion from the end of Section \ref{Swc:3111} we can identify the region where the constant steady state is unstable. This leads to the following restriction for two remaining parameters of the system $\ell$ and $D$:
$$
\frac{\ell}{0.75}<\alpha_c:=\frac{(4-\pi)(\sqrt{2}+1)}{\pi}\quad\mbox{and}\quad
(0.75)^2>\frac{D\pi^2(2+\sqrt{2})}{8\lr{\alpha_c-\frac{\ell}{0.75}}},
$$
which allows to approximate the instability region for this particular case as
$D<D(\ell)=0.1781(0.4948-\ell)$. We also introduce a notation
$\ell_c=R\alpha_c$, which in this case gives $\ell_c=0.4948$. The parameter $\ell_c$ denotes the value of $\ell$ above which  the constant steady state is always stable independently of the value of the parameter $D$.

\begin{figure}[h!]
\includegraphics[scale=0.28]{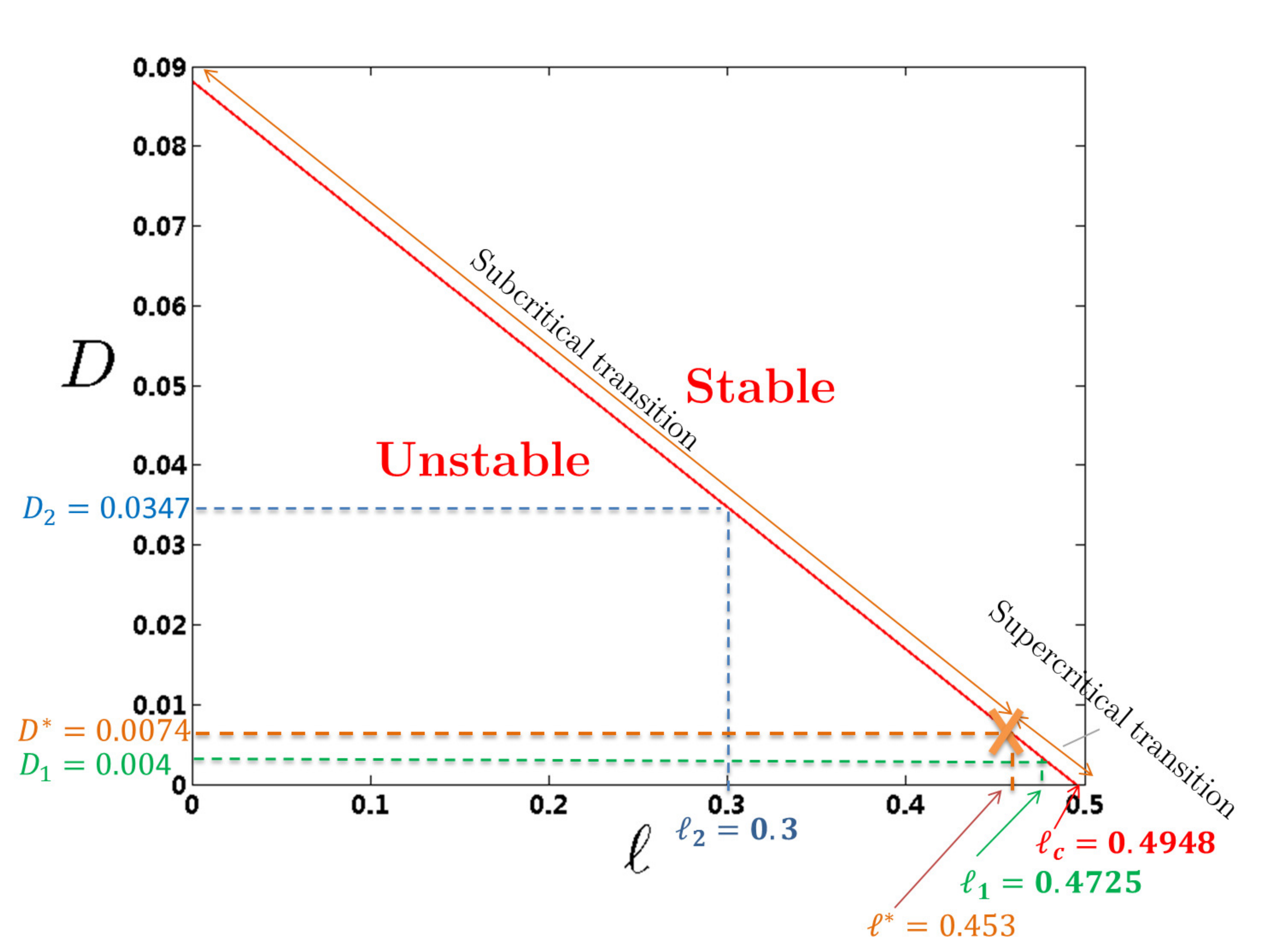}
\caption{Bifurcation diagram in the 1-dimensional case. The critical value for $\ell$, $\ell_c=0.4948$, above which the constant steady state is stable for all values of $D$ is indicated in red. The change of bifurcation type is located at $(\ell^\ast,D^\ast)=(0.4530,0.0074)$ and indicated in orange. For the numerical study, we choose two values of $\ell$: (i) $\ell=\ell_1= 0.4725$, for which a supercritical bifurcation occurs at $D< D_{1}=0.0040$ (indicated in green), and (ii) $\ell=\ell_2= 0.3$, for which a subcritical bifurcation occurs at $D< D_{2}=0.0347$ (indicated in blue). }
\label{Fig1}
\end{figure}

Using~\eqref{Vk} and Proposition \ref{bifurc}, we check that the bifurcation changes its character for $\ell=\ell^\ast$, where $\ell^\ast=0.75\frac{(\pi-4)\sqrt{2}+2}{\pi(\sqrt{2}-1)}\approx 0.4530$. Recall that our criterion did not involve the parameter $D$, therefore the bifurcation is supercritical if only $\ell\in  \lr{\ell^\ast,\ell_c}\approx(0.4530, 0.4948)$, and subcritical if $\ell \in(0,l^\ast)\approx(0, 0.4530)$. 
The value of parameter $D$ corresponding to the instability threshold for $l=l^\ast\approx 0.4530$ is denoted by $D^\ast$ and it is equal to $0.0074$. All of these parameters are presented on the  Fig.~\ref{Fig1}, below.

\subsubsection{Macroscopic model}

We now make use of the numerical scheme developed in \cite{CaChHu} to analyze the macroscopic equation~\eqref{Macro} with the potential~\eqref{TVH} in the unstable regime. The choice of the numerical scheme is due to its free energy decreasing property for equations enjoying a gradient flow structure such as~\eqref{Macro}. Keeping this property of gradient flows is of paramount importance in order to compute the right stationary states in the long time asymptotics. In fact, under a suitable CFL condition the scheme is positivity preserving and well-balanced, i.e., stationary states are preserved exactly by the scheme. 

To check the correctness of the criterion from Proposition \ref{bifurc} we consider two cases corresponding to two different types of bifurcation, as depicted on the Fig. \ref{Fig1}:
\begin{itemize}
\item $\ell_1= 0.4725$ for different values of the noise $D$, where we expect a supercritical (continuous) transition for {$D<D_{1} = 0.0040$};
\item $\ell_2= 0.3$ for different values of the noise $D$, where we expect a subcritical (discontinuous) transition for $D< D_2 = 0.0347$.
\end{itemize}

In order to trace the influence of the diffusion on the type of
bifurcation, for fixed $\ell_1$, $\ell_2$, we will be looking for the
values of diffusion coefficients $D_{1,\lambda}$, $D_{2,\lambda}$
such that
$$
D_{1,\lambda}\uparrow D_{1}=0.0040,\quad D_{2,\lambda}\uparrow D_{2}=0.0347.
$$
Recall that according to  \cite{BDZ2016}, the parameter $\lambda$ defined in~\eqref{lambda1} measures the distance from the instability threshold. We will use this information to determine the values of parameters $D_{1,\lambda}=D_{1,\lambda}(\lambda)$ and $D_{2,\lambda}=D_{2,\lambda}(\lambda)$
computed from \eqref{lambda1}. We consider 14 different values for subcritical and supercritical case, as specified in Table \ref{Macro_table}.

  \begin{table}[h!]
\centering
\begin{tabular}{c|c|c|c}
{}& $\lambda$ & $D_{1,\lambda}$  & $D_{2,\lambda}$ \\
\hline
1& 0.0010& 0.0030
    &    0.0338  \\
2& 0.0009& 0.0031
    &   0.0339  \\
3& 0.0008& 0.0032
    &   0.0340\\
4& 0.0007& 0.0033
    &   0.0340\\
5& 0.0006& 0.0034
    &   0.0341\\
6& 0.0005& 0.0035
    &   0.0342\\
7& 0.0004& 0.0036
    &   0.0343\\
8& 0.0003& 0.0037
    &       0.0344\\
9& 0.0002& 0.0038
    &   0.0345\\
10& 0.0001& 0.0039
    &   0.0346\\
\bf{11}& \bf{0}& \bf{0.0040}
    &   \bf{0.0347} \\
12& -0.0001& 0.0041
    &0.0348\\
13& -0.0002& 0.0042
    &   0.0349\\
14& -0.0003& 0.0043
    &    0.0350\\
\hline
\end{tabular}
\caption{Table of parameters $D_{1,\lambda}$ (supercritical), and $D_{2,\lambda}$ (subcritical) for the numerical simulations in the macroscopic case with highlighted values corresponding to the phase transition. \label{Macro_table}}
\end{table}

Moreover, in \cite{BDZ2016} the authors proved that the perturbation $\rho(t)$ of the constant steady state satisfies the following equation
\begin{equation}
\rho(t,x) = A(t) e_{1}+A^\ast(t) e_{-1} +A^2(t) h_{2} e_{2} + (A^{\ast})^2(t) h_{-2} e_{-2} + O((A,A^\ast)^3),
\label{eq:expg}
\end{equation}
where
\begin{equation}
\dot{A}=\lambda A +8\frac{\pi^4}{L^2}\frac{\hat{V}_{1}}{2\lambda-\lambda_2}\left(2\hat{V}_{2}-\hat{V}_{1} \right)|A|^2A+ O((A,A^\star)^4),
\label{eq:reduced}
\end{equation}
and
$$
h_{2}= -\frac{4\pi^2}{L}\frac{\hat V_1}{(2\lambda-\lambda_2)},\quad h_{-2}= -\frac{4\pi^2}{L}\frac{\hat V_{-1}}{(2\lambda-\lambda_2)}.
$$
Equation~\eqref{eq:reduced} means that for the supercritical bifurcation we can observe a saturation. This means that before stabilizing $A(t)$ first grows exponentially until the r.h.s. of \eqref{eq:reduced} is equal to zero, i.e. for
\eq{\label{mod}
|A|=\frac{\sqrt{\lambda}L}{2\sqrt{2}\pi^2}\sqrt{\frac{2\lambda-\lambda_2}{-\hat V_1(2\hat V_2-\hat V_1)}}.
}
Using  this information to estimate the r.h.s. of~\eqref{eq:expg}, we obtain that
\eq{\label{criterion}
|\rho(t,x)|\approx 2|A|+\frac{\lambda L}{\pi^2(2\hat V_2-\hat V_1)} =
\frac{\sqrt{\lambda}L}{\sqrt{2}\pi^2}\sqrt{\frac{2\lambda-\lambda_2}{-\hat V_1(2\hat V_2-\hat V_1)}}+\frac{\lambda L}{\pi^2(2\hat V_2-\hat V_1)}.
}
This condition gives us the upper estimate for the amplitude of perturbation $\rho$ when the steady state is achieved, that is after the saturation. The derivation of Proposition \ref{bifurc} in \cite{BDZ2016},  assumes sufficiently small perturbation of the steady state. Therefore, the initial data for our numerical simulations should be least smaller than the value of $|A|$ corresponding to the saturation level. It turns out that $|A|$ computed in~\eqref{mod} is always less than $\sqrt{\lambda}$, so the the size of initial perturbation of the steady state should be also taken in this regime. 
If we choose the initial data for the numerical simulations of the supercritical case in this regime, we should see a continuous decay of the saturated amplitude of perturbation to $0$, as $\lambda$ decreases. We will perturb the initial data for the subcritical case similarly, showing that even though the smallness restriction is respected, the saturated amplitude of perturbation is a discontinuous function of $\lambda$.

In what follows, we perturb the constant initial condition by the
first Fourier mode:
$$
f_0(x)=\frac{1}{2L}+{\delta(\lambda)} \cos\lr{\frac{x\pi}{L}},
$$
with ${\delta(\lambda) \leq\sqrt{\lambda}}$. In the numerical
simulations, we consider the case ${\delta}=0.01$. In order to distinguish between the homogeneous steady-states (corresponding to the stable regime) and the aggregated steady-states (corresponding to the unstable regimes), we compute the following quantifier $Q$ on the density profiles of the numerical solutions:
\eq{\label{Q}
Q = \sqrt{c_1^2 + s_1^2},}
\noindent where
$$
c_1 = \frac{1}{L} \int_{-L}^L f(T_{max}, x)
\cos\lr{\frac{x\pi}{L}} dx,\qquad s_1 = \frac{1}{L} \int_{-L}^L
f(T_{max}, x) \sin\lr{\frac{x\pi}{L}} dx,
$$
where $T_{max}$
corresponds to the formation of the steady state. Note that (i) if the steady state is homogeneous in space then $Q=0$, and (ii) if $f$
is a symmetric function with respect to $x$, then $Q = c_1$.

To estimate $T_{max}$ we use the following criterion. From the
theory \cite{CMV03}, we know that steady states are positive
everywhere and the quantity $\xi=D\log\vr+ V\ast\vr$ is equal to
some constant $C$. We then compute the distance of $\xi$ from
its mean value:
$$
{\xi}^{\star}(t)=\max_{x\in[-L,L]}\Big|\xi(t,
x)-\frac{1}{2L}\int_{-L}^L\xi(t, x)\ dx\Big|.
$$
The steady state
is achieved if $\xi^*$ is sufficiently close to 0, and in our
numerical scheme we continue the computations until $t=T_{max}$
for which,
${\xi}^\star(T_{max})<10^{-7}$. The computed values are presented
in the Tables \ref{Table:super} and \ref{Table:sub} in the
Appendix. In Fig. \ref{Fig:lambda}, we show the values of the order parameter $Q$ as a function of the noise intensity $D$ for both types of bifurcation.
\begin{figure}[h!]
\includegraphics[scale = 0.42]{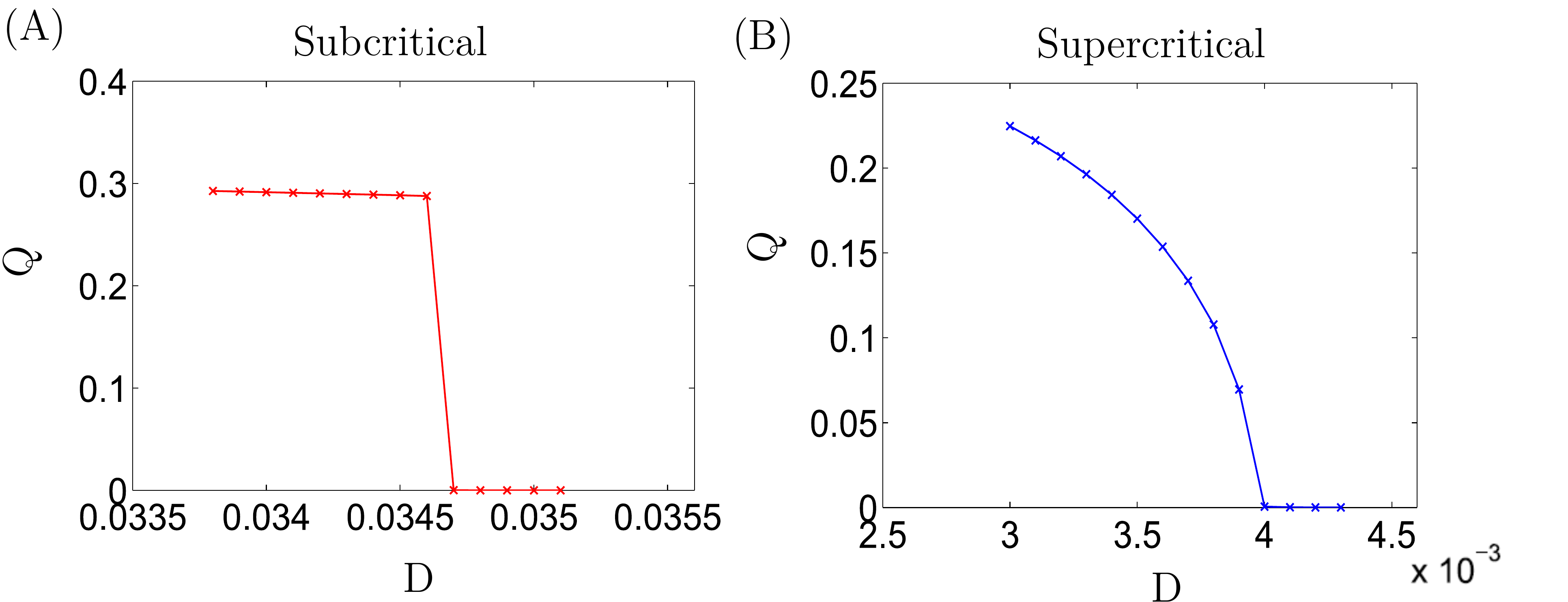}
\caption{Order parameter $Q$ as a function of the diffusion  parameter  $D$ for the macroscopic model for (A) $\ell = 0.3$ (subcritical case) and (B) $\ell=0.4725$ (supercritical case). }
\label{Fig:lambda}
\end{figure}
As shown in Fig. \ref{Fig:lambda}, the quantifier $Q$ indeed undergoes a
discontinuous transition around $D = 0.0347$ for $\ell = 0.3$
(subcritical case, Fig.\ref{Fig:lambda} (A)) and a smooth transition around $D
= 0.004$ for $\ell = 0.4725$ (supercritical case, Fig.\ref{Fig:lambda} (B)).
These results show that the numerical solutions are in very good
agreement with the theoretical predictions.

\begin{figure}[h!]
\includegraphics[scale = 0.45]{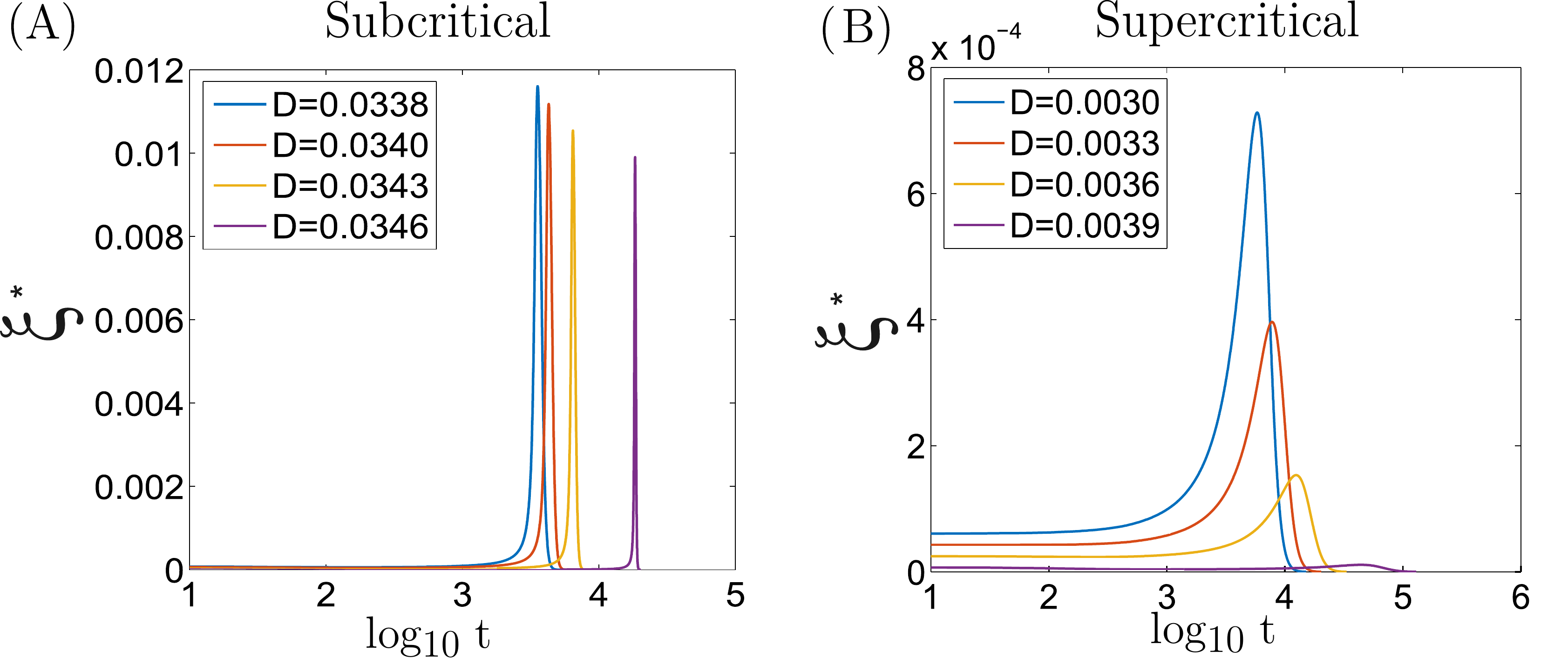}
\caption{Values of $\xi^*$ as a function of $\log_{10}t$ computed on the steady-states of the macroscopic model for (A): $\ell=0.3$ (subcritical case) and (B) $\ell=0.4725$ (supercritical case).}
\label{Fig:xi1D}
\end{figure}

In order to check the accuracy of our prediction of the value of $T_{max}$, we show in Fig.~\ref{Fig:xi1D} the graph of $\xi^*(t)$ for several values of $D$ in the supercritical and the subcritical cases (see Table \ref{Macro_table}). As shown by Fig.~\ref{Fig:xi1D}, we observe a very sharp change of $\xi^*$ for the subcritical bifurcation and much smoother one for the supercritical case.
The amplitude change of $\xi^*$ is also a good indication of the type of bifurcation. As for the order parameter, we see that for the subcritical bifurcation it is on similar level (Fig. \ref{Fig:xi1D} (A)) for all values of $D$, while for the supercritical bifurcation it decays to $0$ (Fig. \ref{Fig:xi1D} (B)). We will use this observation to analyze the results of the 2-dimensional simulations later on.

\begin{figure}
\begin{floatrow}
\ffigbox{%
 \includegraphics[scale = 0.5]{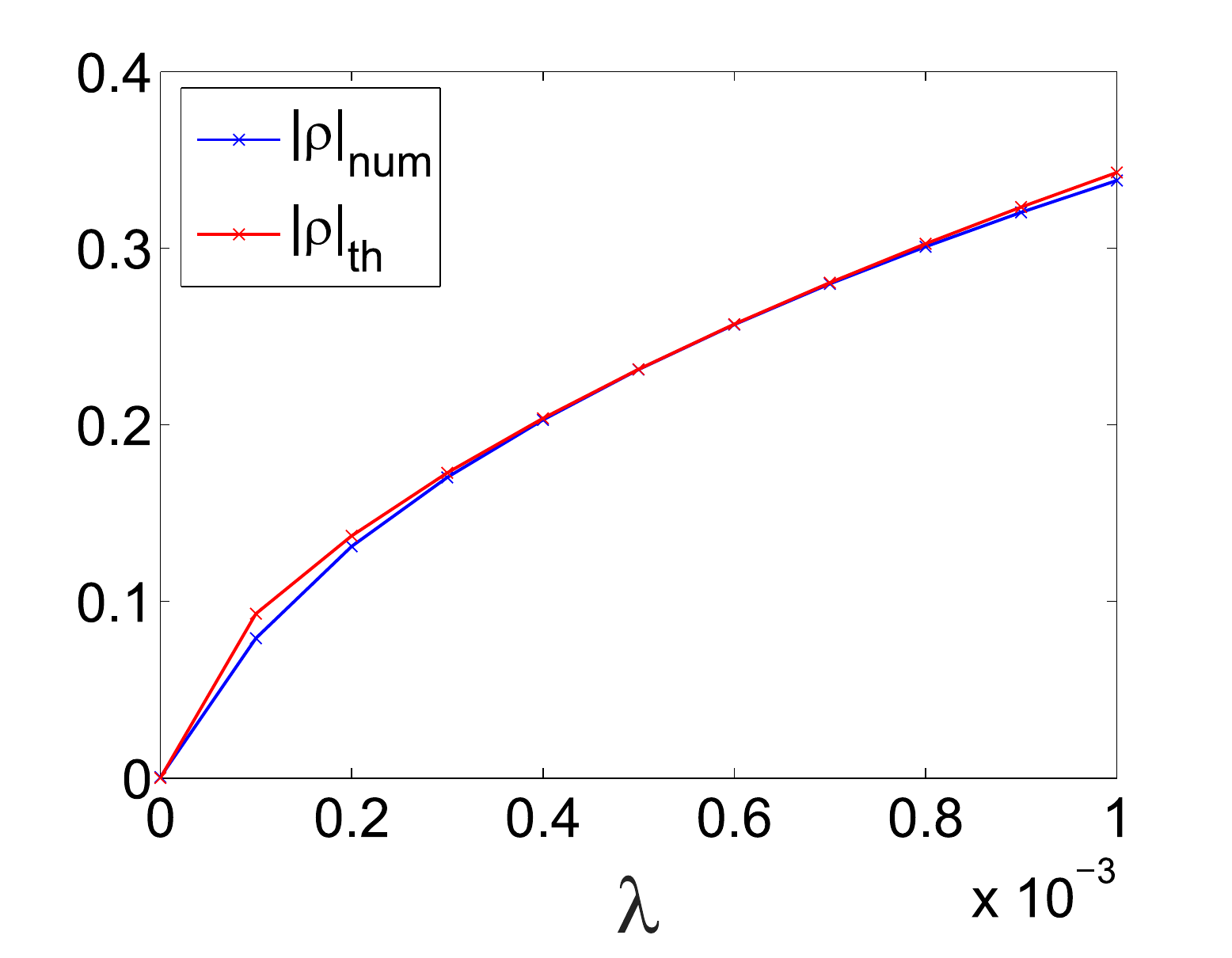}%
}{%
  \caption{Comparison of theoretical $|\rho|_{th}$ with the numerical $|\rho|_{num}$.}\label{Qrho}%
}

\capbtabbox{%
 \begin{tabular}{c|c|c|c}
 $\lambda$  &$|\rho|_{num}$ & $|A|$&$|\rho|_{th}$\\
 \hline
0.0010  &   0.3384 &  0.1094& 0.3428
\\
 0.0009     &0.3203 & 0.1058& 0.3233
\\
 0.0008     &0.3008 & 0.1017& 0.3025
\\
0.0007      & 0.2797 & 0.0968& 0.2805
\\
 0.0006     & 0.2567 & 0.0912& 0.2569
\\
0.0005      &0.2312 & 0.0847& 0.2314
\\
0.0004      &0.2028 & 0.0770& 0.2036
\\
0.0003      &0.1701 & 0.0678&  0.1727
\\
 0.0002     &0.1311 & 0.0562& 0.1371
\\
 0.0001     &0.0790 & 0.0403& 0.0930
\\
 0      &0.0005 & 0& 0
\\
 \hline
\end{tabular}}{%
  \caption{Theoretical ($|\rho|_{th}$) vs numerical ($|\rho|_{num}$) values for the size of perturbation}\label{Table:Qrho}%
}
\end{floatrow}
\end{figure}

Finally, we can also check how the theoretical prediction of the size of
perturbation from \eqref{criterion}  is confirmed by our numerical
results. For this purpose, we compute the maximum of the
perturbation once the steady state is achieved:
$$
|\rho|_{num}=\|f(T_{max},x)-\vr^\star\|_{L^\infty((-L,L))}
$$
for all the points of supercritical bifurcation. The results are
presented on Figure \ref{Qrho} and in the Table \ref{Table:Qrho}.

We now aim to perform the same stability analysis on the microscopic model from the Section \ref{SSec:micro} -- the starting point of the derivation of the macroscopic model.

\subsubsection{Microscopic model}

Here, we aim to perform simulations of the microscopic model from Section \ref{SSec:micro}, rescaled with the scaling from the Section \ref{Ss:scaling}. After rescaling and if we consider an explicit Euler scheme in time (see Appendix \ref{App:MicroNumerics}), we can show that Eq. \eqref{Micro_Eq} between time steps $t^n$ and $t^n+\Delta t^n$ reads (in non-dimensionalized variables):
\eq{\label{Micro}
X_i^{n+1} = X_i^n - \nabla_{X_i} W(X^n) \Delta t^n + \sqrt{2{D} \Delta t^n},
}
\noindent where $W$ is defined by \eqref{W}. Between two time steps, new links are created between close enough pairs of particles {\bf that are not already linked} with probability $\mathbb{P}_f = 1 - e^{{\nu_f \Delta t^n}/{((N-1)\varepsilon^2)}}$ and the existing links disappear with probability $\mathbb{P}_d = 1 - e^{-{\nu_d \Delta t^n}/{((N-1)\varepsilon^2)}}$. Therefore, the rescaled version of the microscopic model features a very fast link creation/destruction rate, as the linking and unlinking frequencies are supposed to be of order $1/\varepsilon^2$, for small $\varepsilon$. Note also that to capture the right time scale, the time step $\Delta t$ must be decreased with $\varepsilon$, which makes the microscopic model computationally costly for small values of $\varepsilon$. For computation time reasons, we also consider the limiting case $\varepsilon = 0$ of the microscopic model; we can show that it reads:
\eq{\label{Micro_eps0}
X_i^{n+1} = X_i^n - \nabla_{X_i} W_0(X^n) \Delta t^n + \sqrt{2{D} \Delta t^n},
}
\noindent where
$$
W_0(X) = \sum_{i,j |\  |X_i - X_j|\leq R} V(X_i,X_j).
$$
Note that in this regime, no fiber link remains and particles interact with all of their close neighbours. The limit $N \rightarrow \infty$ of this limiting microscopic model should exactly correspond to the macroscopic model~\eqref{Macro_sys} (see for instance \cite{BCC11,FHM12,CCH,GQ15} for studies of mean-field limits including, as in the present case, singular forces). If not otherwise stated, the values of the parameters in the microscopic simulations are given by Table~\ref{Micro_table}.
  \begin{table}[h!]
\centering
\begin{tabular}{c|c|c}
Parameter & Value & Interpretation\\
\hline
$L$ & 3 & Domain half size\\
$\delta$ & 0.1 & Maximal step\\
$T_f$ & 20 & Final simulation time\\
$\xi_{init}$ & 0.1 & Initial fraction $\frac{K}{N}$\\
$\nu_d$ & $1$ & Unlinking frequency\\
$\nu_f$ & $1$ & Linking frequency\\
$R$ & 0.75 & Detection radius for creation of links\\
$\ell$ & adapted & Spring equilibrium length\\
$\kappa$ & {2} & Spring force between linked fibers\\
$D$ & adapted & Noise intensity\\
$\varepsilon$ & adapted & Scaling parameter\\
\hline
\end{tabular}
\caption{Table of parameters (non dimensionalized values) \label{Micro_table}}
\end{table}

As for the macroscopic model, the order of the particle system at equilibrium is measured by the quantifier $Q$ defined by Eq. \eqref{Q}, where the integrals are computed using the trapezoidal rule. To compute the density of agents $f(x)$ in the microscopic simulations, we divide the computational domain $[-L,L]$ into $N_x$ boxes of centers  $x_i$ and sizes $dx = \frac{L}{N_x}$ and for $i= 1\hdots N_x$, we estimate
$${f_i} = \frac{N_i}{2 N L},$$
\noindent where $f_i=f(x_i)$ and $N_i$ are respectively the density and the number of agents whose centers belong to the interval $[-L + (i-1)dx, -L + i\, dx]$. {Following the analysis of the macroscopic model, we explore the same two cases:
 $\ell_1= 0.4725$, $D_{1}=0.0040$, and $\ell_2= 0.3$, $D_{2}=0.0347$ to check whether they correspond to the super and subcritical bifurcations, respectively.

In Fig.~\ref{Micro_transition_eps0}, we show the values of $Q$ plotted as functions of the noise intensity $D$ computed from the simulations of the scaled microscopic model~\eqref{Micro} at equilibrium, for two different values of $\ell$: $\ell = 0.3$ (A), $\ell = 0.4725$ (B), and different values of $\varepsilon$: $\varepsilon = \frac{1}{6}$ (blue curves), $\varepsilon = \frac{1}{8}$ (orange curves), $\varepsilon = \frac{1}{12}$ (black curves), and the limiting case $"\varepsilon = 0"$ (Eq. \eqref{Micro_eps0}, green curves). For each $\ell$, we superimpose the values of $Q$ obtained with the simulations of the macroscopic model (red curves). As expected, we observe subcritical transitions for $\ell = 0.3$, and a supercritical transition for $\ell = 0.4725$. As $\varepsilon$ decreases, the values of the noise intensity $D$ for which the transitions occur get closer to the theoretical values predicted by the analysis of the macroscopic model. These results show that the scaled microscopic model has the same properties as the macroscopic one, and that the values of the parameters ($\ell, D$) which correspond to a bifurcation in the steady states tend, as $\varepsilon \rightarrow 0$, to the ones predicted by the analysis of the macroscopic model. Indeed for the limiting case $"\varepsilon=0"$ of the microscopic model, we obtain a very good agreement between the micro- and macro- formulations showing that the microscopic model behaves as predicted by the analysis of the macroscopic model.

\begin{figure}[h!]
\centering
\includegraphics[scale = 0.5]{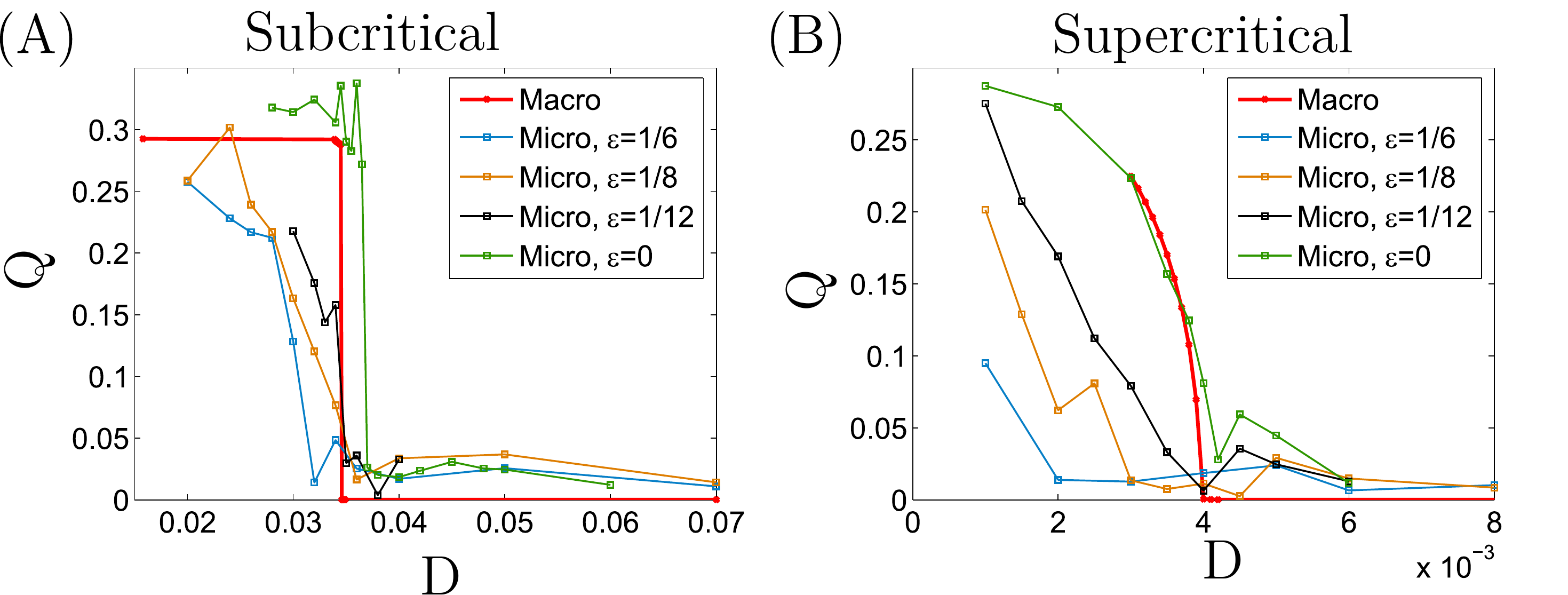}
\caption{Values of $Q$ plotted as function of the noise intensity $D$ computed from the numerical solutions at equilibrium of the macroscopic model (red curves), and of the microscopic model for $\varepsilon=1/16$ (blue curves), $\varepsilon=1/8$ (orange curves), $\varepsilon=1/12$ (black curves) and limiting case "$\varepsilon = 0$" (Eq.~\eqref{Micro_eps0}, green curves). (A) For $\ell = 0.3$ (subcritical bifurcation), (B) for $\ell = 0.4725$ (supercritical bifurcation). For small $\varepsilon$ and these two values of $\ell$, we recover the bifurcation types predicted by the analysis of the macroscopic model. As $\varepsilon$ decreases, the critical values of $D$ for which the transitions occur get closer to the ones of the macroscopic model, and in the limiting case '$\varepsilon=0$' in the microscopic model, we obtain a very good agreement between the microscopic and macroscopic models. \label{Micro_transition_eps0}}
\end{figure}

 It is noteworthy that the small differences observed in the values of the transitional $D$ (subcritical case, Fig.\ref{Micro_transition_eps0} (A)) can be due to the fact that we use a finite number of $N=1000$ particles for the microscopic simulations, whereas the macroscopic model is in the limit $N\rightarrow \infty$. However, these differences are very small when we consider the limit case $\varepsilon=0$ for the microscopic model. Indeed, the relative error between the microscopic and macroscopic transitional $D$, $\frac{|D_{mic} - D_{mac}|}{D_{mac}}$ is $7\%$ for $\ell = 0.3$, and $5\%$ for $\ell = 0.4725$.

We now aim to compare the profiles of the solutions between the
microscopic and macroscopic models, to numerically validate the
derivation of the macroscopic model from the microscopic dynamics.

\subsubsection{Comparison of the density profiles in the microscopic and macroscopic models}

Here, we aim to compare the profiles of the particle densities of
the microscopic model with the ones of the macroscopic model as
functions of time. As shown in the previous section, for $\varepsilon$ small enough, we recover the bifurcation and bifurcation types observed from the macroscopic model with the microscopic formulation, with very good quantitative agreement when considering the limiting microscopic
model~\eqref{Micro_eps0} with '$\varepsilon = 0$'. The simulations of the microscopic model are very time
consuming for small values of $\varepsilon$, because we are obliged
to consider very small time steps. Here, due to 
computational time constraints, we therefore compare the results of the
macroscopic model~\eqref{Macro_sys} with
$\varepsilon=0$ for which the time step can be taken much larger and independent of $\varepsilon$.

In order to have the same initial condition for both the
microscopic and macroscopic models, we initially choose the
particle positions for both models such that:
$$
f_0(x) = \frac{1}{2L} + \delta(\lambda) \cos \frac{x\pi}{L}.
$$
We send the reader to Appendix \ref{App:MicroNumerics} for the numerical method used to set the initial conditions of the microscopic model. Because of the stochastic nature of the model, the microscopic model 
does not preserve the symmetry of the solution,
contrary to the macroscopic model (where noise results in a deterministic diffusion
term). To enable the comparison between the macroscopic and
microscopic models, we therefore re-center the periodic domain of
the microscopic model such that the center of mass of the
particles is located at $x=0$ (center of the domain). To this aim,
given the set of particles $X_j, j=1 \hdots N$, we reposition all the particles at points $\tilde{X}_j, j=1\hdots N$ such that:
$$
\tilde{X}_j = \begin{cases}
X_j - X_m \hspace{2.6cm} \text{if } |X_j - X_m|\leq L\\
 X_j - X_m - 2L \frac{X_j - X_m}{|X_j - X_m|} \quad \text{if } |X_j - X_m|> L,
\end{cases}
$$
\noindent where $X_m$ is the center of mass computed on a periodic domain:
$$
X_m =\frac{L}{\pi} \; arg \bigg(\frac{1}{2}\sum_{j=1}^N e^{\frac{i \pi X_j}{L}} \bigg).
$$
\noindent Finally, in order to decrease the noise in the data of the microscopic simulations due to the random processes, the density of particles is computed on a set of several simulations of the microscopic model.

In Fig.~\ref{MicMacl47}, we show the density distributions of the macroscopic model (continuous lines) and of the microscopic one with "$\varepsilon = 0$" (circle markers) at different times, for $\ell = 0.4725$ and $\ell = 0.3$ respectively. For each value of $\ell$, we consider two values for the noise intensity $D$: For $\ell = 0.4725$ we study the cases $D = 0.003$ and $D=0.0003$, and for $\ell = 0.3$ we choose $D = 0.0338$ and $D = 0.0034$. Note that all these values are in the unstable regime.

\begin{figure}[h!]
\centering
\includegraphics[scale = 0.47]{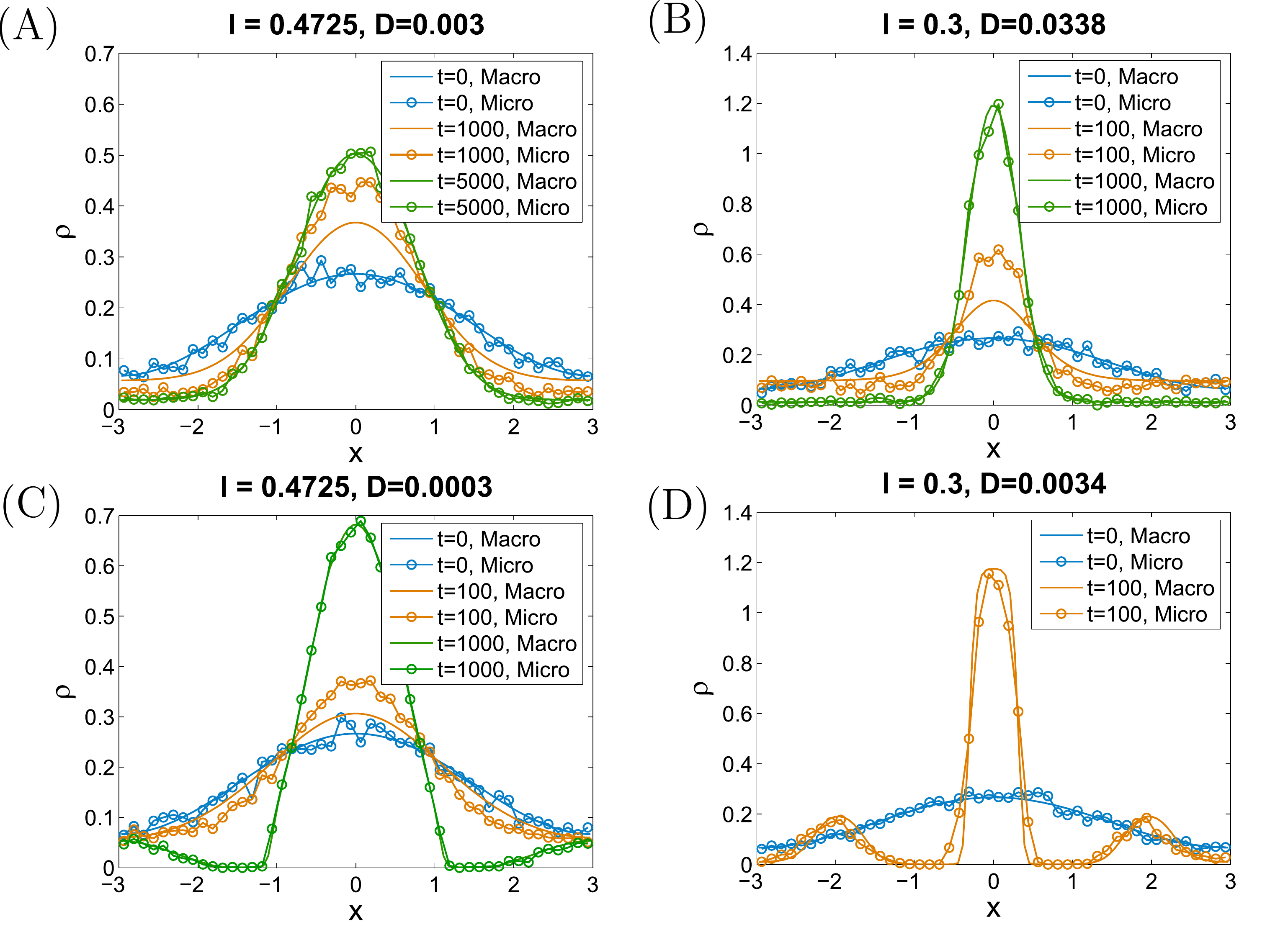}
\caption{Comparison of the density distributions between the macroscopic model and the microscopic one with "$\varepsilon = 0$", for different times and two values of $\ell$: $\ell = 0.4725$ (A and C), $\ell = 0.3$ (B and D).  Continuous lines: solution of the macroscopic model, with circles: solution of the microscopic model  with $\varepsilon = 0$, averaged over 6 simulations. For each value of $\ell$, we consider two different noise intensities $D$: for $\ell = 0.4725$ we use $D = 0.003$ (A) and $D=0.0003$ (C), and for $\ell=0.3$ we use $D=0.0338$ (B) and $D=0.0034$ (D). \label{MicMacl47}}
\end{figure}

As shown by Figs.~\ref{MicMacl47}, we obtain a very good agreement between the solutions of the macroscopic model and of the microscopic one with "$\varepsilon = 0$". Close to the transitional $D$ (Fig. \ref{MicMacl47} (A) and (B)), the particle density converges in time towards a Gaussian-like distribution  for both the microscopic and macroscopic models. Note that the microscopic simulations seem to converge in time towards the steady state faster than the macroscopic model (compare the orange curves on the top panels). This change in speed can be due to the fact that the microscopic model features finite number of particles while the macroscopic model is obtained in the limit of infinite number of particles. Therefore, in the macroscopic setting, each particle interacts with many more particles than in the microscopic model, which could result in a delay in the aggregation process.

When far from the transitional $D$ in the unstable regime (Fig. \ref{MicMacl47} (C) and (D)), one can observe the production of several bumps in the steady state of the particle density. The production of several particle clusters in these regimes shows that the noise triggers particle aggregation. For small noise intensity, local particle aggregates are formed which fail to detect neighboring aggregates. As a result, one can observe several clusters in the steady state, for small enough noise intensities. These bumps are observed for both the microscopic and macroscopic models, showing again a good agreement between the two dynamics.

In the next section, we present a numerical study of the
macroscopic model in the 2-dimensional case. As mentioned previously, the
microscopic model is in very good agreement with the macroscopic
dynamics for small values of $\varepsilon$ as in the 1-dimensional case. Its
simulations are, however, very time consuming, due to the need of
very small time steps. As a result, the microscopic model is not
suited for the study of very large systems such as the ones
considered in the 2-dimensional case. We therefore provide a numerical
2-dimensional study using the macroscopic model only.

%
%

\section{Analysis of the macroscopic model in the 2-dimensional case}\label{Sec:macro2D}
\subsection{Theoretical results}
In this section, we first recall some theoretical results from
\cite{BDZ2016} for the two-dimensional periodic domain. We will
focus on the square periodic domain $[-L,L]\times[-L,L]$, since
the rectangular case can be, in agreement with the analysis in
\cite{BDZ2016}, reduced to the one-dimensional case studied above.

The starting point for the phase transition analysis is the linearized equation
$$
\partial_{t}  \rho={D\lap_{x}  \rho}+\rho^*\lap (V\ast \rho),
$$
in which the spatially homogeneous distribution $\rho^*$ is now equal to $\frac{1}{(2L)^2}$. Applying the Fourier transform to this equation, we obtain
$$
\pt\hat \rho_{k_1,k_2}=-{\pi^2}\frac{k_1^2+k_2^2}{L^2} \lr{D+\hat{
V}_{k_1,k_2}}\hat \rho_{k_1,k_2}:=\lambda_{k_1,k_2}\hat
\rho_{k_1,k_2}
$$
and we denote $\lambda_{\pm1,0}=\lambda_{0,\pm1}=\lambda$. The Fourier transform of the potential $V$ is given by
\eq{\label{fourV}
\hat{ V}_{k_1,k_2}=\frac{\pi}{L^2}\Big(\frac{\pi R^3 l}{2z_{k_1,k_2}^2}\big[J_1(z_{k_1,k_2})H_0(z_{k_1,k_2})-J_0(z_{k_1,k_2})H_1(z_{k_1,k_2})\big]-\frac{R^4}{z_{k_1,k_2}^2}J_2(z_{k_1,k_2})\Big),
}
where we denoted
$$
z_{k_1,k_2}=\frac{\pi R}{L}\sqrt{{k_1^2+k_2^2}},
$$
$J_i$ are Bessel function of order~$i$ \eqh{
J_i(x)=\sum_{m=0}^\infty \frac{(-1)^m}{m!
\Gamma(m+1+i)}\lr{\frac{x}{2}}^{2m+i}, } and $H_i$ are the Struve
functions defined by \eqh{
H_i(x)=\sum_{m=0}^\infty\frac{(-1)^m}{\Gamma(m+3/2)\Gamma(m+i+3/2)}\lr{\frac{x}{2}}^{2m+i+1}.
} Again, fixing the ratio $\frac{R}{L}\leq1$, the relation between
$D,l$ and $R$ for the phase transition can be read from the
condition $\lambda=0$, which yields \eq{\label{DV10} D+\hat
V_{1,0}=0,} which due to~\eqref{fourV} gives \eqh{
D\pi+R^2\lr{\frac{\pi}{2}\frac{l}{R}\lr{J_1(z_{1,0})H_0(z_{1,0})-J_0(z_{1,0})H_1(z_{1,0})}-J_2(z_{1,0})}=0.
} The relevant criterion for the type of bifurcation in the
two-dimensional case then reads:

\begin{prop}\label{bifurc_2D}
Assume $D$ is varied such that it crosses the bifurcation point \eqref{fourV}, and such that $\lambda_{k_1,k_2}$ remains negative for all 
 $k_1,k_2$ such that $|k_1|+|k_2|>1$, let 
$$
c= \frac{\hat{V}_{1,0}(2\hat{V}_{2,0}-\hat{V}_{-1,0})}{D+\hat
V_{2,0}},\quad d=-\left| 4
\frac{\hat{V}_{1,0}\hat{V}_{1,1}}{D+\hat V_{1,1}}\right|,
$$
then, 
\begin{itemize}
\item if $c<d$,
the constant steady state exhibits a supercritical bifurcation,
\item if $c>d$,
the constant steady state exhibits a subcritical bifurcation.
\end{itemize}
\end{prop}
Note that in the two-dimensional case, the bifurcation criterion involves also parameter $D$. On the other hand, the instability threshold $D$ is given as a function of $\alpha$, and can be calculated using~\eqref{DV10}.

\subsection{Numerical results}

We first compute the approximate instability regime for the following three cases:

\indent 1. For $R/L=1$, $z_{1,0}=\pi$, the constant steady state is unstable for
$$
\frac{\ell}{R}<\alpha_c:=0.6620\quad \mbox{and}\quad D< 0.2334
R^2{\lr{\alpha_c-\frac{\ell}{R}}}.
$$
\indent 2. For $R/L=1/2$, $z_{1,0}=\frac{\pi}{2}$, the constant steady
state is unstable for
$$
\frac{\ell}{R}<\alpha_c:=0.7333\quad \mbox{and}\quad D< 0.1084
R^2{\lr{\alpha_c-\frac{\ell}{R}}}.
$$
\indent 3. For $R/L=1/4$, $z_{1,0}=\frac{\pi}{4}$, the constant steady
state is unstable for
$$
\frac{\ell}{R}<\alpha_c:=0.7462\quad \mbox{and}\quad D< 0.0312
R^2{\lr{\alpha_c-\frac{\ell}{R}}}.
$$

Therefore, for $L=3$, the  criterion from Proposition \ref{bifurc_2D} gives the following outcomes:
\begin{enumerate}
\item For $R/L=1$, the steady state exhibits a supercritical bifurcation for $\alpha\in(0.1016, 0.5818)$, and a subcritical bifurcation for $\alpha\in(0,0.1016)\cup(0.5818,0.6620)$.

\item For $R/L=1/2$, the steady state exhibits only a subcritical bifurcation.

\item For $R/L=1/4$, the steady state exhibits only a subcritical bifurcation.
\end{enumerate}
\begin{figure}[h!]
\includegraphics[scale = 0.4]{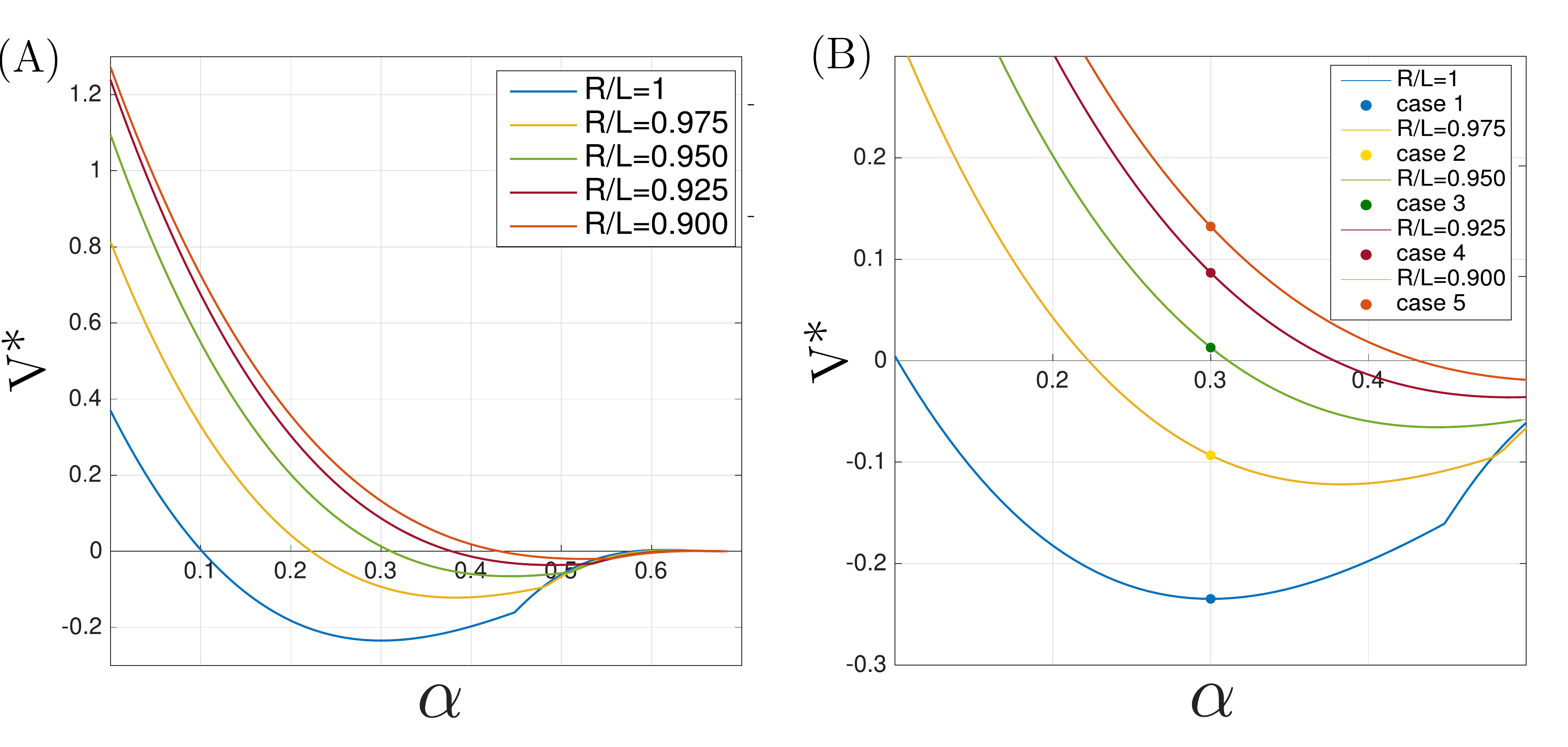}
\caption{(A) Values of $V^*$ as a function of $\alpha$ for different ratios $R/L$. (B) Zoom on the values of $V^*$ around $\alpha=0.3$ for different ratios $R/L$, with marker points for $V^*$ at $\alpha=0.3$ (used in the numerical simulations).} \label{Fig:xi}
\end{figure}

This computation confirms the theoretical prediction from
\cite{BDZ2016} that the  smaller $\frac{R}{L}$ is, the more likely it is that the bifurcation is of the subcritical type. The
different types of bifurcation happen only for $\frac{R}{L}$ close
to $1$, otherwise the bifurcation is always subcritical. To
understand this behavior, we compute
$$V^*=
 {\hat{V}_{1,0}(2\hat{V}_{2,0}-\hat{V}_{-1,0})}\left|{D+\hat V_{1,1}}\right|+4
 \lr{D+\hat V_{2,0}} \left| {\hat{V}_{1,0}\hat{V}_{1,1}}\right|.
$$
From Proposition \ref{bifurc_2D} it follows that if $V^*>0$ the
bifurcation is subcritical, otherwise it is supercritical. We
depict the function $V^*(\alpha)$, where $\alpha=\frac{\ell}{R}$ for
different values of $\frac{R}{L}\in [0.9,1]$ on Fig.
\ref{Fig:xi} (A). We see that decreasing the ratio $\frac{R}{L}$ causes that more and more of the graph of $V^*(\alpha)$ lies above $0$. This means that for most of the values of $\alpha\in [0,\alpha_c]$ the bifurcation is subcritical.

We will study all of the five cases  from Figure \ref{Fig:xi}
(B) corresponding to different values of $\frac{R}{L}$ but the same
value of $\alpha=\frac{\ell}{R}=0.3$. The theoretical prediction is that the
first two cases $\frac{R}{L}=1$ and $\frac{R}{L}=0.975$ correspond to a
supercritical (continuous) bifurcation while the cases 3-5
correspond to the subcritical (discontinuous) bifurcation.

We perturb the constant initial data as in the 1-dimensional case, namely we take
$$
f_0(x,y)=\frac{1}{4L^2}+\delta \cos\lr{\frac{x\pi}{L}},
$$
with $\delta=0.01$, and similarly to the 1-dimensional case we compute the
value of the order parameter $Q$
$$Q=\frac{1}{2L^2}\int_{-L}^{L}\int_{-L}^{L}f(T_{max},x,y)\cos\lr{\frac{x\pi}{L}}\, dx\, dy,$$
where we used the empirical observation that  the steady state is always  symmetric with respect to $(x,y)=(0,0)$.
For the stopping time criterion we take the same as in 1-dimensional
case, namely ${\xi}^\star(T_{max})<10^{-7}$.

In Fig. \ref{Fig:lambda2}, we show the values of the order
parameter $Q$ as function of the noise intensity $D$ for both
types of bifurcation for cases 1 and 5, based on the Tables
\ref{T2D_1} and  \ref{T2D_5} from the Appendix.
\begin{figure}[h!]
\includegraphics[scale = 0.4]{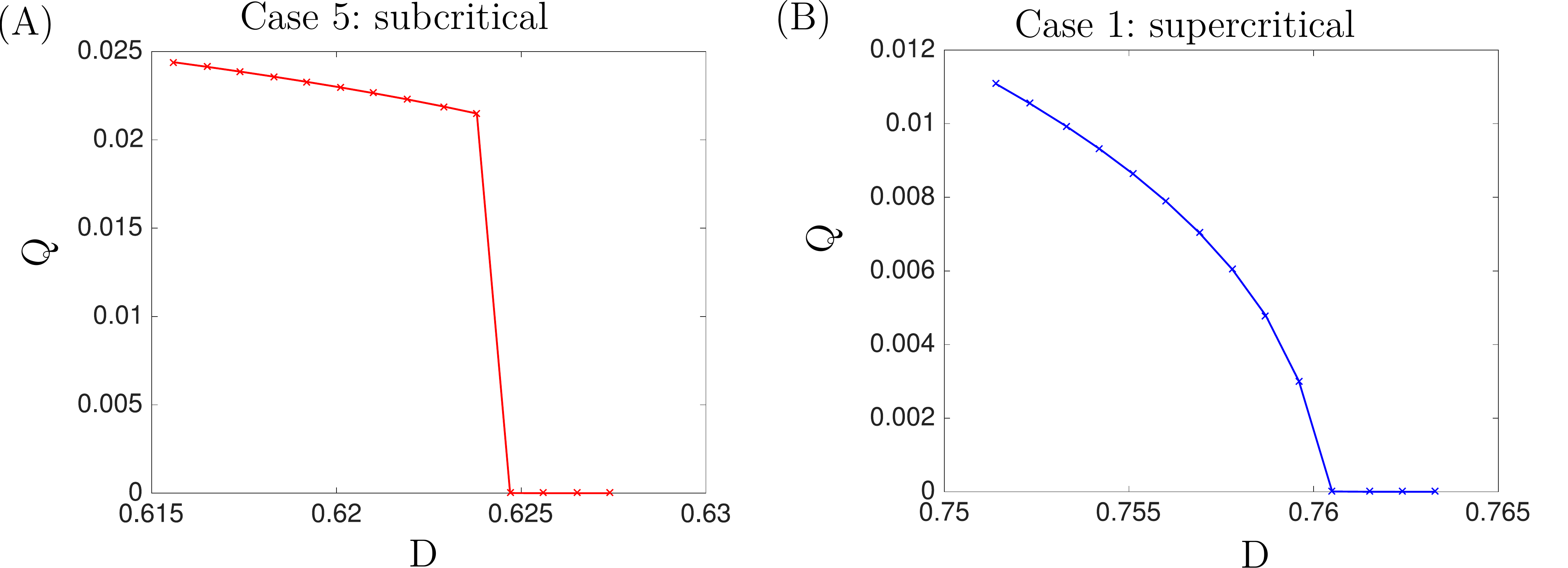}
\caption{Quantifier $Q$ as a function of the diffusion coefficient $D$ computed on the steady states of the macroscopic 2-dimensional model. These bifurcation diagrams have been generated from the data of Tables
\ref{T2D_1} and  \ref{T2D_5} from the Appendix \ref{Ab2}.}
\label{Fig:lambda2}
\end{figure}
As shown by Fig. \ref{Fig:lambda2}, we indeed obtain a
supercritical (continuous) transition in the values of $Q$ as
function of the noise $D$ in case 1 (Fig. \ref{Fig:lambda2} (B)), while the transition is discontinuous (subcritical) in
case 5 (Fig. \ref{Fig:lambda2} (A)). These results therefore show
that the numerical results are in good agreement with the
theoretical predictions and provide a validation of the numerical 
approximation and simulations of the macroscopic model.

The difference between the bifurcation types is also reflected in the amplitude of the steady state. For both types of bifurcation, i.e. for cases 1 and cases 5 we plot the final steady states on Fig. \ref{Fig:rho2}. The density profile for the supercritical bifurcation (Fig. \ref{Fig:rho2} (B)) is much lower and rounded than the one for the subcritical bifurcation (Fig. \ref{Fig:rho2} (A)).
\begin{figure}[h!]
\includegraphics[scale = 0.42]{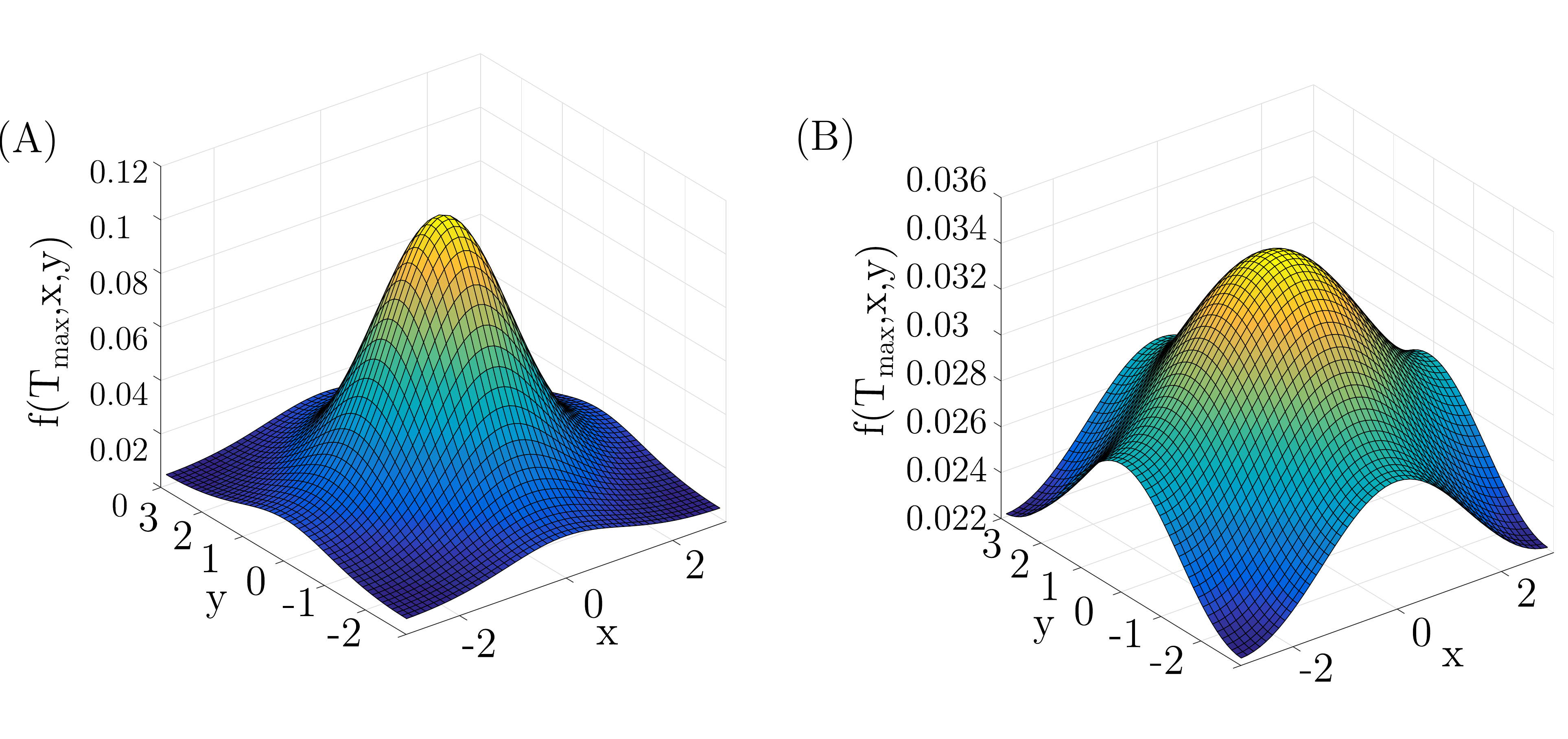}
\caption{Final density profile in the subcritical case 5 for $D_5=0.6238$ (A) and in the supercritical case 1 for $D_1=0.7596$(B). The values of parameters $D_1$ and $D_5$ in both cases correspond to $\lambda=0.001$ in Table \ref{Macro2_table}.}
\label{Fig:rho2}
\end{figure}

Moreover, as in the 1-dimensional case, we can check that the different bifurcation diagrams correspond to different shapes of $\xi^*$. Below on Figs. \ref{Fig:super12} and \ref{Fig:sub345}, we present the graphs of $\xi^*(t)$ for all five cases depicted at Fig. \ref{Fig:xi} (B). For each of the cases we present the graph of $\xi^*(t)$ for five different values of diffusion parameter $D$ as specified in Table \ref{Macro2_table} below.
  \begin{table}[h!]
\centering
\begin{tabular}{c|c|c|c|c|c|c}
{}& $\lambda$ & $D_{1,\lambda}$  & $D_{2,\lambda}$& $D_{3,\lambda}$& $D_{4,\lambda}$ & $D_{5,\lambda}$  \\
\hline
1& 0.005& 0.7560
    &   0.7254& 0.6923& 0.6570& 0.6201\\
2& 0.004& 0.7569
    &   0.7264& 0.6932 & 0.6579& 0.6210\\
3& 0.003& 0.7578 
    &       0.7273& 0.6941 & 0.6588 & 0.6219 \\
4& 0.002& 0.7587
    &   0.7282& 0.6950& 0.6597& 0.6229\\
5& 0.001& 0.7596
    &   0.7291& 0.6959 & 0.6607& 0.6238\\
\hline
\end{tabular}
\caption{Table of parameters $D_{1,\lambda}$ and $D_{2,\lambda}$ (supercritical), $D_{3,\lambda}$, $D_{4,\lambda}$, and$D_{5,\lambda}$ (subcritical) for the numerical simulations in 2-dimensional case. \label{Macro2_table}}
\end{table}
\begin{figure}[h!]
\includegraphics[scale = 0.4]{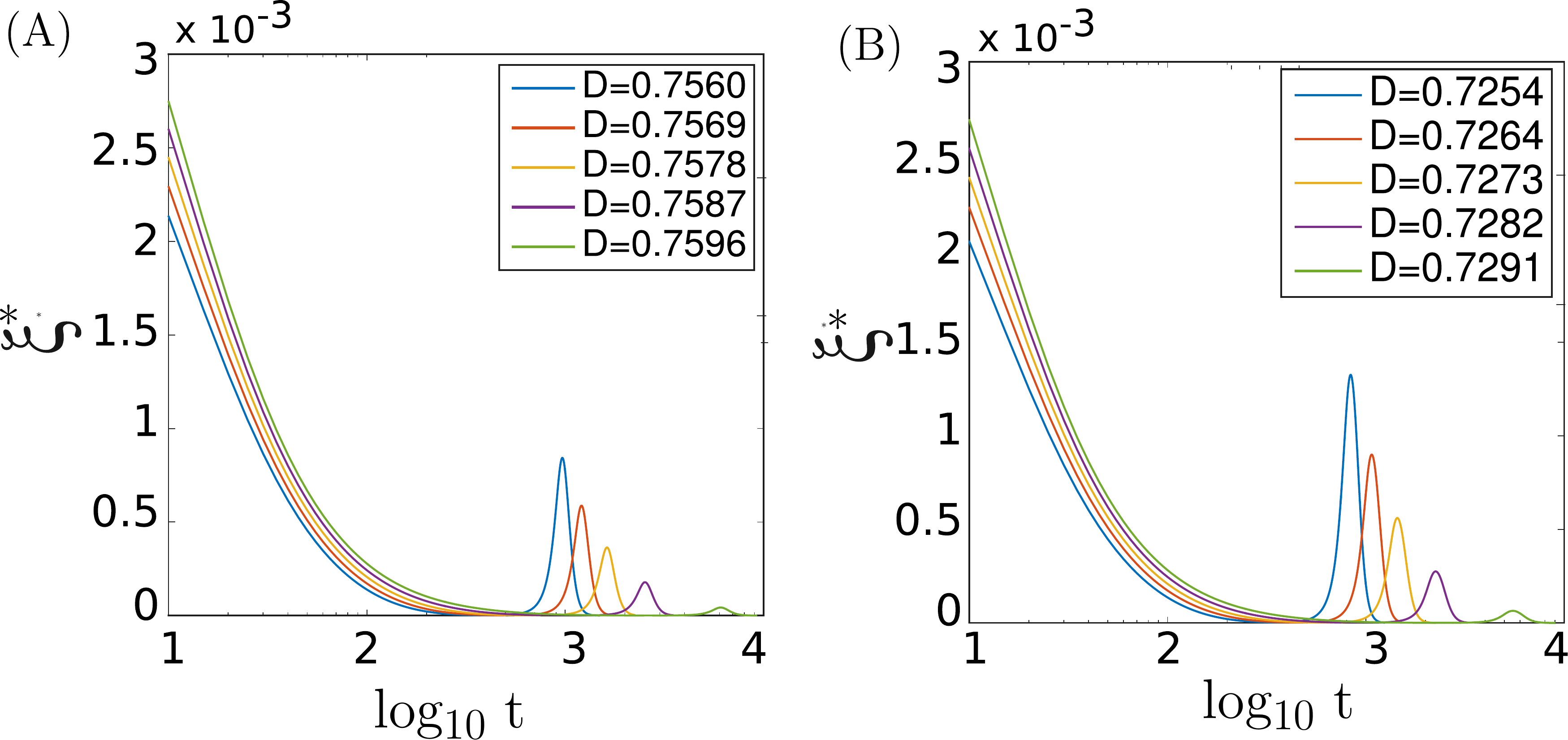}
\caption{Values of $\xi^*(\log_{10}t)$ for the supercritical bifurcation  for (A) case 1 and (B) case 2.}
\label{Fig:super12}
\end{figure}
\begin{figure}[h!]
\includegraphics[scale = 0.4]{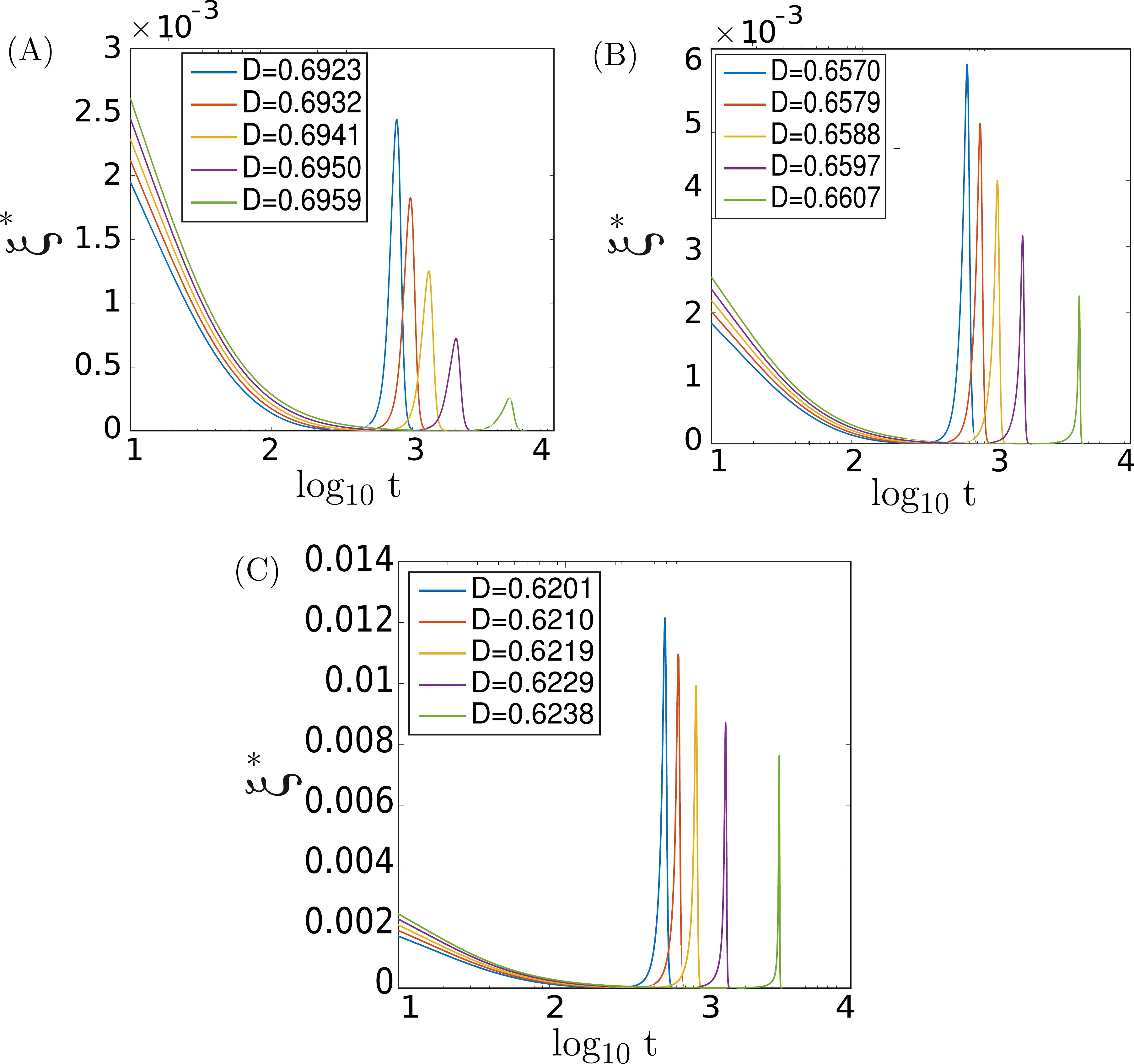}
\caption{Values of $\xi^*(\log_{10}t)$ for the subcritical bifurcation for (A) case 3, (B) case 4 and (C) case 5.}
\label{Fig:sub345}
\end{figure}
As shown on Fig. \ref{Fig:super12}, the graph of
$\xi^*(t)$ undergoes smooth changes for the different values of
the noise $D$, highlighting a bifurcation of supercritical type. Fig. \ref{Fig:sub345} shows that $\xi^*(t)$
undergoes sharp changes for the different values of the
noise $D$, highlighting the subcritical type of bifurcation, as
predicted by the theoretical analysis of the macroscopic model in
the 2-dimensional case. Close to the transition zone (case 3,  $\frac{R}{L} = 0.95$,
Figure \ref{Fig:sub345} (A)), the changes in $\xi^*$ are smoother than
for smaller values of $\frac{R}{L}$ (Fig. \ref{Fig:sub345} (B) and (C)), but the transition is still subcritical as can be confirmed by the values of order parameter $Q$ given in Table \ref{T2D_3}.

\section{Conclusion}

In this paper, we have provided a numerical study of a macroscopic model derived from an agent-based formulation for particles
interacting through a dynamical network of links. In the 1-dimensional case, we were first able to recover numerically the subcritical and
supercritical transitions undergone by the steady states of the macroscopic model, in the regime predicted by the theoretical
nonlinear analysis of the continuous model. Moreover, the numerical simulations of the rescaled microscopic model revealed
the same bifurcations and bifurcation types as obtained with the macroscopic model, with very good precision as $\varepsilon$ goes to zero in the microscopic setting. Finally, when considering the limiting case '$\varepsilon=0$' in the microscopic model, we obtained a very good agreement between the profiles of the solutions of the micro- and macro- models. It is noteworthy that both models also feature the same dynamics in time, with a slight delay in the macroscopic simulations compared to the microscopic dynamics. This delay may be due to the fact that the microscopic simulations are performed with a finite number of particles while the macroscopic model is in the limit of infinite number of individuals. However, as for very small values of $\varepsilon$ the
simulations of the microscopic dynamics are very time consuming, we were not able to extend the numerical study to a higher number
of particles.

For the sake of completeness, we finally presented numerical simulations of the model in the 2-dimensional case. For computational reasons, we were not able to perform 2-dimensional simulations of the
microscopic model, and we chose to focus on the macroscopic model.
In the 2-dimensional case, we were once again able to numerically recover
supercritical and subcritical transitions in the steady states, as
function of the noise intensity $D$, in the same regime as
predicted by the theoretical analysis of the macroscopic model.
These results validate the theoretical analysis, the numerical
method and the simulations developed for the macroscopic model.

By providing a numerical comparison between the micro- and macro- dynamics, this study shows that the macroscopic model considered
in this paper is indeed a relevant tool to model particles interacting through a dynamical network of links. As a main
advantage compared to the microscopic formulation, the macroscopic
model enables to explore large systems with low computational cost
(such as 2-dimensional studies), and is therefore believed to be a powerful
tool to study network systems on the large scale. Direct
perspectives of these works include the derivation of the
macroscopic model in a regime of non-instantaneous
linking-unlinking of particles. The hope is to understand deeper
how the local forces generated by the links are expressed at the
macroscopic level. The model could be improved by taking into
account other phenomena such as external forces, particle
creation/destruction etc.  Finally, rigorously proving the
derivation of the macroscopic model from the particle dynamics
will be the subject of the future research.

\appendix
\section{Appendix--On the micro model}\label{App:MicroNumerics}

The scaling of Section 2.2 obliges us to increase the length of the domain when decreasing $\varepsilon$. For convenience, we rather work with fixing the domain $\Omega$. We therefore use the following scaling:

$\tilde{x} = x$, $f^\varepsilon = f, g^\varepsilon = g$, $\tilde{\kappa} = \varepsilon^{-1} \kappa$, $\tilde{D} = \varepsilon D$, $\tilde{R} = R$, $\tilde{l} = l$, $\tilde{t} = \varepsilon t$, $\tilde{\nu}_f = \varepsilon^2 \nu_f$, $\tilde{\nu}_d = \varepsilon^2 \nu_d$.

One can check that this scaling, after letting $\ep\to 0$ leads to the same macroscopic system~\eqref{Macro_sys}. The microscopic model then reads (dropping the tildes for clarity purposes):
$$
X_i^{n+1} = X_i^n - \nabla_{X_i} W(X^n) \Delta t^n + \sqrt{2{D} \Delta t^n}.
$$
\noindent Between two time steps, new links are created between close enough pairs of particles {\bf that are not already linked} with probability $\mathbb{P}_f = 1 - e^{-\nu_f^N \Delta t^n}$ and new links disappear with probability $\mathbb{P}_d = 1 - e^{-\nu_d^N \Delta t^n}$, where we denoted $\nu_f^N = \tilde{\nu}_f/((N-1)\varepsilon^2)$, $\nu_d^N = \tilde{\nu}_d/(N-1)\varepsilon^2)$. Here, the time step $\Delta t^n$ is chosen such that the particle motion is bounded by the numerical parameter $\delta>0$ and such that $\nu_f^N \Delta t^n<0.1$, $\nu_d^N \Delta t^n <0.1$ (to capture the right time scale). To this purpose, we set:
$$
\Delta t^n = \min\lr{\frac{\delta}{N_{lpf} R \kappa}\, ,\ \frac{0.1}{\max(\nu_f^N,\nu_d^N)}},
$$
\noindent where $N_{plf}$ is the maximal number of links per fiber. Note that as the links are dynamical $N_{plf}$ might change during the course of the simulation, making the time step dependent on the current step. For the particle simulations, we suppose that the number of particles is large enough so that we can set $\nu_f^N = \frac{\nu_f}{N-1}$ and $\nu_d^N = \nu_d$. Finally as explained in the main text, we initially choose the particle positions for both models such that:
$$
f_0(x) = \frac{1}{2L} + \delta(\lambda) \cos \frac{x\pi}{L}.
$$
\noindent For the microscopic model, the initial positions of the particles are set such that, given a random position $X \in [-L,L]$ and a random number $\eta_2 \in [-1,1]$, we let the probability of creating a new particle at position $X+ \eta_2 \Delta x - \frac{\Delta x}{2}$ be $\frac{\Delta x}{2L} + \delta(\lambda) \Delta x \cos \frac{\pi X}{L}$.

\section{Appendix--Tables}
\subsection{Numerical results for the 1-dimensional case}

  \begin{table}[h!]
\centering
\begin{tabular}{c|c|c|c|c}
$l_1=0.4725$& $D_{1,\lambda}$ & $\lambda$ & $T_{max}$  &$Q$\\
\hline
super-1& 0.0030 & 0.0010
    & 1.52e+4     &0.2247      \\
super-2& 0.0031 & 0.0009
    &1.65e+4        &0.2163 \\
super-3& 0.0032 & 0.0008
    &1.82e+4        &0.2070 \\
super-4& 0.0033 & 0.0007
    & 2.04e+4       &0.1963 \\
super-5& 0.0034 & 0.0006
    &2.38e+4    &0.1842 \\
super-6& 0.0035 & 0.0005
    &2.74e+4        &0.1702 \\
super-7& 0.0036 & 0.0004
    &3.34e+4        &0.1537 \\
super-8& 0.0037 & 0.0003
    &4.35e+4     & 0.1336       \\
super-9& 0.0038 & 0.0002
    &6.38e+4        & 0.1078    \\
super-10& 0.0039    & 0.0001
    &1.30e+5        & 0.0696    \\
super-11& 0.0040    & 0
    &1.63e+4         &5.0e-4    \\
super-12& 0.0041    & -0.0001
    &5.99e+4            & 1.3e-4\\
super-13& 0.0042    & -0.0002
    &3.86e+4        &7.2e-5
\\
super-14& 0.0043    & -0.0003
    &2.90e+4     & 5.0e-5   \\
    \hline
\end{tabular}
\caption{The supercritical bifurcation in the 1-dimensional case.} \label{Table:super}
\end{table}
  \begin{table}[h!]
\centering
\begin{tabular}{c|c|c|c| c}
$l_2=0.3$& $D_{2,\lambda}$ & $\lambda$ & $T_{max}$  &$Q$\\
 \hline
sub-1& 0.0338& 0.0010 & 0.54e+3     &0.2926\\
sub-2& 0.0339   & 0.0009
    &0.57e+4        &0.2921\\
sub-3& 0.0340   & 0.0008
    &0.61e+4        &0.2915\\
sub-4& 0.0340   & 0.0007
    &0.61e+4        &0.2915\\
sub-5& 0.0341   & 0.0006
    &0.67e+4        &0.2909\\
sub-6& 0.0342   & 0.0005
    &0.74e+4        &0.2903\\
sub-7& 0.0343   & 0.0004
    &0.83e+4        &0.2897\\
sub-8& 0.0344   & 0.0003
    &0.98e+4        &0.2891\\
sub-9& 0.0345   & 0.0002
    &1.26e+4        &0.2884\\
sub-10& 0.0346  & 0.0001
    &2.03e+4        &0.2878\\
sub-11& 0.0347  & 0
     &1.01e+5     &2.2e-4 \\
sub-12& 0.0348  & -0.0001
    &4.98e+4        &9.5e-5\\
sub-13& 0.0349  & -0.0002
    &3.43e+4        &6.1e-5\\
sub-14& 0.0350  & -0.0003
    &2.66e+4        &4.5e-5\\
    \hline
\end{tabular}
\caption{The subcritical bifurcation in the 1-dimensional case.} \label{Table:sub}
\end{table}

\subsection{Numerical results for the 2-dimesional case}\label{Ab2}

  \begin{table}[h!]
\centering
\begin{tabular}{c|c|c|c|c}
case 1& $D_{\lambda}$ & $\lambda$ & $T_{max}$ &$Q$ \\
 \hline
 super-1&0.7514&0.010&791.1&0.0111\\
 super-2&0.7523&0.009&871.5&0.0106\\
 super-3&0.7533&0.008&984.9&0.0099\\
 super-4&0.7542&0.007&1.1183e+3&0.0093\\
 super-5&0.7551&0.006&1.2971e+3&0.0086\\
super-6& 0.7560 & 0.005
    &       1.5498e+3 & 0.0079\\
super-7& 0.7569 & 0.004
    & 1.9324e+3 &0.0070\\
super-8& 0.7578 & 0.003
    &   2.5875e+3   &0.0060     \\
super-9& 0.7587 & 0.002
    &  3.9821e+3& 0.0048    \\
super-10& 0.7596    & 0.001
    &  9.2876e+3&0.0030     \\
super-11& 0.7605&0.000&7.7450e+3&8.3456e-6\\
super-12&0.7615&-0.001&2.6994e+3&2.0849e-6\\
super-13&0.7624&-0.002&1.7778e+3&1.2446e-6\\
super-14&0.7633&-0.003&1.3437e+3&8.8705e-7\\
\hline
\end{tabular}
\caption{The supercritical bifurcation in the 2-dimensional case: $\alpha=0.3$, $R/L=1$.}\label{T2D_1}
\end{table}
  \begin{table}[h!]
\centering
\begin{tabular}{c|c|c|c|c}
case 2& $D_{\lambda}$ & $\lambda$ & $T_{max}$ &$Q$ \\
 \hline
super-1& 0.7254 & 0.005
    &       1.3932e+3& 0.0102\\
super-2& 0.7264 & 0.004
    & 1.8138e+3&0.0091\\
super-3& 0.7273 & 0.003
    &   2.5098e+3   &0.0078     \\
super-4& 0.7282 & 0.002
    &  4.1365e+3& 0.0062    \\
super-5& 0.7291 & 0.001
    &  9.0104e+3&0.0042 \\
\hline
\end{tabular}
\caption{The supercritical bifurcation in the 2-dimensional case: $\alpha=0.3$, $R/L=0.975$.}\label{T2D_2}
\end{table}
  \begin{table}[h!]
\centering
\begin{tabular}{c|c|c|c|c}
case 3& $D_{\lambda}$ & $\lambda$ & $T_{max}$ &$Q$ \\
 \hline
sub-1& 0.6923   & 0.005
    &       1.1447e+3&      0.0142\\
sub-2& 0.6932   & 0.004
    & 1.4385+3& 0.0133  \\
sub-3& 0.6941   & 0.003
    &   1.9591e+3   &0.0123     \\
sub-4& 0.6950   & 0.002
    &  3.1522e+3& 0.0109    \\
sub-5& 0.6959   & 0.001
    &  6.4408e+3 &0.0093        \\
\hline
\end{tabular}
\caption{The subcritical bifurcation in the 2-dimensional case: $\alpha=0.3$ $R/L=0.950$.}\label{T2D_3}
\end{table}
  \begin{table}[h!]
\centering
\begin{tabular}{c|c|c|c|c}
case 4 & $D_{\lambda}$ & $\lambda$ & $T_{max}$ &$Q$ \\
 \hline
sub-1& 0.6570   & 0.005
    &       858.4&     0.0190\\
sub-2& 0.6579   & 0.004
    & 1.0410e+3&0.0185  \\
sub-3& 0.6588   & 0.003
    &   1.3428e+3   &0.0179     \\
sub-4& 0.6597   & 0.002
    &  1.9528e+3& 0.0173    \\
sub-5& 0.6607   & 0.001
    &  4.6192e+3 &0.0165        \\
\hline
\end{tabular}
\caption{The subcritical bifurcation in the 2-dimensional case:  $\alpha=0.3$ $R/L=0.925$.}\label{T2D_4}
\end{table}
  \begin{table}[h!]
\centering
\begin{tabular}{c|c|c|c|c}
case 5 & $D_{\lambda}$ & $\lambda$ & $T_{max}$ &$Q$ \\
 \hline
 sub-1&0.6156&0.010&405.3&0.0244\\
 sub-2&0.6165&0.009&437.8&0.0241\\
 sub-3&0.6174&0.008&478.4&0.0239\\
 sub-4&  0.6183&0.007&530.3&0.0236\\
 sub-5& 0.6192&0.006&598.7&0.0233\\
sub-6& 0.6201   & 0.005
    &       693.1 & 0.0230\\
sub-7& 0.6210   & 0.004
    & 832.6& 0.0226 \\
sub-8& 0.6219   & 0.003
    &   1.0628e+3   &0.0223     \\
sub-9& 0.6229   & 0.002
    &  1.6089e+3&    0.0219 \\
sub-10& 0.6238  & 0.001
    &  3.4882e+3&0.0215     \\
sub-11& 0.6247 &0.000&8.0028e+3&8.4646e-6\\
sub-12&0.6256&-0.001&2.9256e+3&2.2626e-6\\
sub-13&0.6265&-0.002&1.8733e+3&1.3058e-6\\
sub-14&0.6274&-0.003&1.3983e+3&9.1773e-7\\
\hline
\end{tabular}
\caption{The subcritical bifurcation in the 2-dimensional case:  $\alpha=0.3$ $R/L=0.900$.}\label{T2D_5}
\end{table}

\section*{Acknowledgments}
JAC was partially supported by the Royal Society and the Wolfson Foundation through a Royal Society Wolfson 
Research Merit Award and by the National Science Foundation (NSF) under grant no. RNMS11-07444 (KI-Net). PD acknowledges support by the Engineering and Physical Sciences 
Research Council (EPSRC) under grants no. EP/M006883/1, EP/N014529/1 and EP/P013651/1, by the Royal Society and the Wolfson Foundation through a Royal Society Wolfson 
Research Merit Award no. WM130048 and by the National Science 
Foundation (NSF) under grant no. RNMS11-07444 (KI-Net). PD is on leave 
from CNRS, Institut de Math\'ematiques de Toulouse, France. DP acknowledges support by the Vienna Science and Technology fund. Vienna project number LS13/029. The work of EZ has been supported by the Polish Ministry of Science and Higher Education grant "Iuventus Plus"  no. 0888/IP3/2016/74. 

\section*{Data availability} No new data were collected in the course of this research.

\end{document}